\documentclass{amsart}

\usepackage{amsmath,amsthm,amssymb}
\usepackage{xcolor}
\usepackage{graphicx}
\usepackage{subcaption}
\usepackage{algorithm}
\usepackage{algorithmicx}
\usepackage{algpseudocode}
\usepackage{enumitem}
\usepackage{mathtools}
\usepackage{comment}
\usepackage{pgf, tikz}
\usetikzlibrary{arrows, automata, positioning}

\newtheorem{definition}{Definition}

\newtheorem{proposition}{Proposition}
\newtheorem{theorem}{Theorem}
\newtheorem{remark}{Remark}

\newtheorem{property}{Property}
\newtheorem{lemma}{Lemma}
\newtheorem{corollary}{Corollary}

\newtheorem{note}{Note}

\theoremstyle{definition}
\newtheorem*{remark*}{Remark}

\title{Flip Paths Between Lattice Triangulations}
\author{William Sims$^1$}
\thanks{$^1$CISE Department, University of Florida, partially supported by NSF DMS 1564480, NSF DMS 1563234, and DARPA HR00111720031.}
\author{Meera Sitharam$^{1,2}$}
\thanks{$^2$Affiliate in Department of Mathematics}
\date{}

\begin{document}
\maketitle

\begin{abstract}
    We present a $O(n^{\frac{3}{2}})$-time algorithm for the \emph{shortest (diagonal) flip path problem} for \emph{lattice} triangulations with $n$ points,  
    improving over previous $O(n^2)$-time algorithms.  For a large, natural class of inputs,  our bound is tight in the sense that our algorithm runs in time linear in the number of flips in the output flip path.  
    Our results rely on an independently interesting structural elucidation of shortest flip paths as the linear orderings of a unique partially ordered set, called a \emph{minimum flip plan}, constructed by a novel use of Farey sequences from elementary number theory.
    Flip paths between general (not necessarily lattice) triangulations have been studied in the combinatorial setting for nearly a century.  In the Euclidean geometric setting, finding a shortest flip path between two triangulations is NP-complete. 
    However, for lattice triangulations, which are studied as spin systems, there are known $O\left(n^2\right)$-time algorithms to find shortest flip paths.
    These algorithms, as well as ours, apply to \emph{constrained} flip paths that ensure a set of \emph{constraint} edges are present in every triangulation along the path.  
    Implications for determining simultaneously flippable edges, i.e. finding optimal simultaneous flip paths between lattice triangulations, and for counting lattice triangulations are discussed.  
\end{abstract}

\section{Introduction and Motivation}
\label{sec:introduction}
\begin{figure}[tbhp]
        \centering
        \begin{subfigure}[t]{0.14\textwidth}
            \centering
            \includegraphics[width=1.05\linewidth]{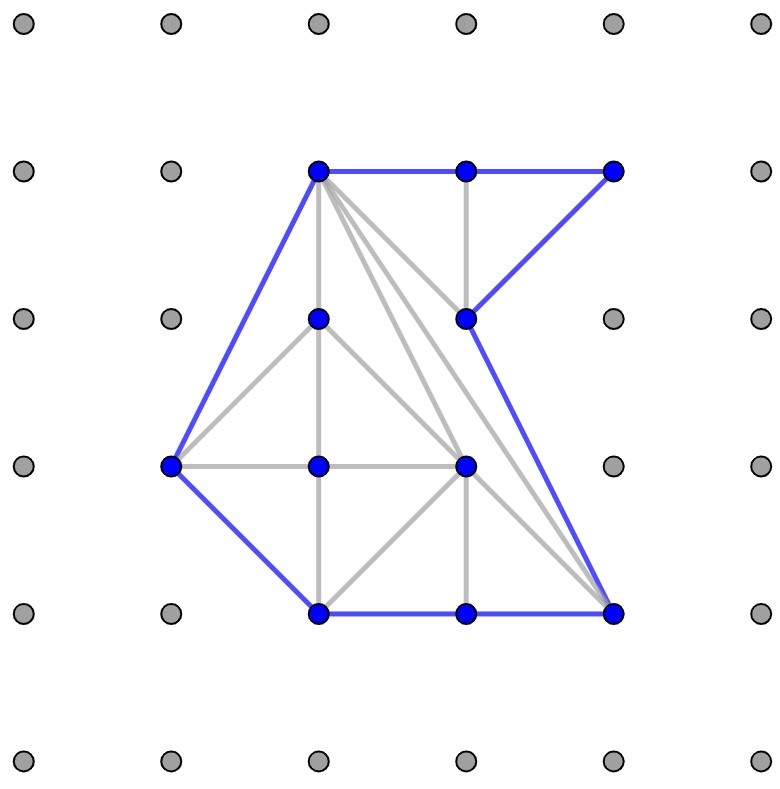}
            \caption{ }
            \label{fig:int_lat}
        \end{subfigure}
        \begin{subfigure}[t]{0.14\textwidth}
            \centering
            \includegraphics[width=0.9\linewidth]{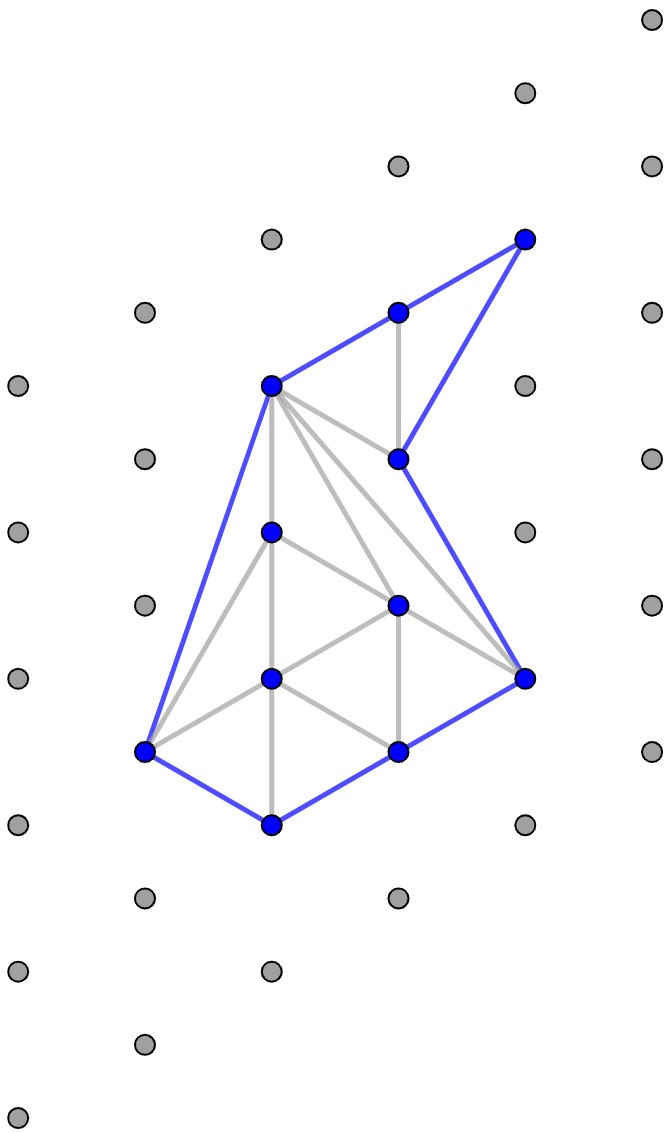}
            \caption{ }
            \label{fig:aff_intlat}
        \end{subfigure}
        \begin{subfigure}[t]{0.13\textwidth}
            \centering
            \includegraphics[width=\linewidth]{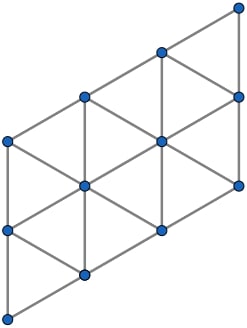}
            \caption{ }
            \label{fig:sec1_fp1}
        \end{subfigure}
        \begin{subfigure}[t]{0.13\textwidth}
            \centering
            \includegraphics[width=\linewidth]{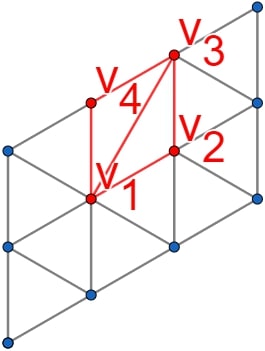}
            \caption{ }
            \label{fig:flip}
        \end{subfigure}
        \begin{subfigure}[t]{0.13\textwidth}
            \centering
            \includegraphics[width=\linewidth]{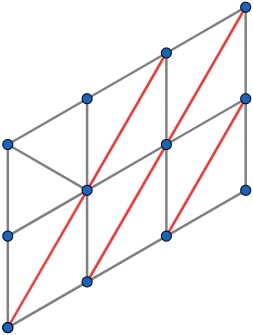}
            \caption{ }
            \label{fig:sec1_fp3}
        \end{subfigure}
        \begin{subfigure}[t]{0.13\textwidth}
            \centering
            \includegraphics[width=\linewidth]{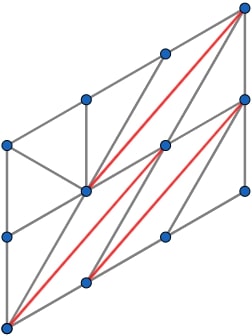}
            \caption{ }
            \label{fig:sec1_fp4}
        \end{subfigure}
        \begin{subfigure}[t]{0.13\textwidth}
            \centering
            \includegraphics[width=\linewidth]{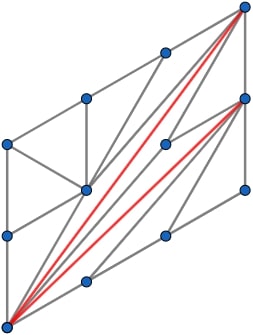}
            \caption{ }
            \label{fig:sec1_fp5}
        \end{subfigure}
        \caption{All figures show triangulations of a point-set bounded by simple, closed polygons.  
        (a) The integer lattice and a lattice triangulation (see below).  
        (b) An affine transformation of (a), yielding the equilateral lattice and the transformed triangulation.  
        (c) - (g) Triangulations along a flip path (Definition \ref{def:flip_flippath}) between the triangulations in (c) and (g).  
        (d) The first flip along this flip path, performed on the red quadrilateral, which replaces the edge $\left(v_2,v_4\right)$ with the edge $\left(v_1,v_3\right)$.  
        See Sections \ref{sec:intro_lat} and \ref{sec:contributions}.}
        \label{fig:lattice_poly_flip}
    \end{figure}

    Given a set of points $P$ in $\mathbb{R}^2$ and a simple, closed polygonal region $\Omega$ with vertices in $P$, a \emph{triangulation} of the point-set $P \cap \Omega$ is an embedding of a graph into $\mathbb{R}^2$ with vertex set $P \cap \Omega$ and non-intersecting straight-line edges that contain exactly two points in $P \cap \Omega$, including the boundary edges of $\Omega$, such that each face of $\Omega$ is a triangle.  
    Note that $\Omega$ need not be convex (see also Section \ref{section:conclusion}).  
    See Figure \ref{fig:lattice_poly_flip} for examples.  
    Throughout this paper we let $S=P \cap \Omega$ and refer to \emph{triangulations of the point-set $S$}.  
    When $S$ is clear from context, we just refer to triangulations.

    \begin{definition}[Diagonal Flip and Flip Path]
        \label{def:flip_flippath}
        For some convex quadrilateral in a triangulation that is formed by two triangles that share an edge, a \emph{(diagonal) flip} is an operation that exchanges the diagonals of the quadrilateral (Figure \ref{fig:flip}).  
        Thus, a flip transforms one triangulation into another.  
        
        A \emph{flip path} is a \emph{starting} triangulation along with a sequence of flips (equivalently, the corresponding sequence of triangulations) ending with a \emph{target} triangulation (Figures \ref{fig:sec1_fp1} - \ref{fig:sec1_fp5}).
    \end{definition}
	
    Flip paths between triangulations have been studied in the combinatorial setting for nearly a century.  
    Here, a \emph{triangulation of a surface} is a simple graph that can be embedded on the surface such that each face is a triangle and any two faces share at most one edge.  
    For some $4$-cycle of the triangulation that contains a single chordal edge, a \emph{combinatorial flip} replaces this chordal edge with the other chordal pair of vertices of the $4$-cycle.  
    We say that there exists a \emph{combinatorial flip path} between two triangulations $T$ and $T'$ of a surface if there exists a sequence of combinatorial flips that transforms $T$ into $T'$.
	
    In the combinatorial setting, Wagner \cite{wagner1936bemerkungen} was the first to show that there exists a combinatorial flip path between any two triangulations of the plane, with the same number of vertices.  
    This result was extended to triangulations of the torus \cite{dewdney1973wagner}, the projective plane and Klein bottle \cite{negami1990diagonal}, and, for triangulations whose graphs have sufficiently many vertices, closed surfaces in general \cite{negami1994diagonal}.  
    See \cite{negami1999diagonal} for a survey of combinatorial flip paths between triangulations of surfaces.
    
    Of interest in distance geometry is a simple constructive proof of the Koebe-Andreev-Thurston circle packing theorem \cite{connelly2020packing} that relies on the existence of a combinatorial flip path between any two triangulations of the plane.
    More generally, length and other structural properties of \emph{shortest flip paths} - i.e., flip paths of minimum length - are metrics on the space of triangulations.  
    
    Returning to our geometric setting in the plane, consider a point-set $S = P \cap \Omega$.  
    Lawson \cite{lawson1972transforming} was the first to prove the existence a flip path between any two triangulations of $S$ when the bounding polygon $\Omega$ is simple, closed, and convex.  
    An extension of this result to the case where $\Omega$ is simple and closed, but not necessarily convex, is given in \cite{dyn1993transforming}, and is attributed to Edelsbrunner.  
    Alternative proofs of the previous two results using only simple geometric properties of polygons can be found in \cite{osherovich2008all}.  
    See \cite{de2010triangulations} for a broad treatment of flip paths between triangulations of a point-set.
    
    Apart from existence results, some work has gone into finding a shortest flip path between two triangulations.  
    This problem was shown to be NP-hard even for triangulations of a point-set consisting only of the vertices of a simple and closed polygon \cite{aichholzer2015flip}.  
    Lawson \cite{lawson1972transforming} gave an algorithm to find \emph{some} flip path between any two triangulations of a point-set $S$ that contains $O(n^2)$ flips, where $n=|S|$. 
    This bound was later proven to be tight \cite{hurtado1999flipping}.  
    
    Finally, in both the combinatorial \cite{galtier2003simultaneous,souvaine2011simultaneously} and geometric \cite{bose2007simultaneous,de2021transforming} settings, properties of \emph{simultaneous flips} - i.e., operations that perform multiple flips at once - and optimal simultaneous flip paths have also been studied.  
    
    Our focus in this paper is  the shortest flip path problem when the desired flip path is between \emph{lattice triangulations} 
    - i.e. triangulations of a \emph{lattice point-set} $S=L \cap \Omega$, where $L$ is an affine transformation of the integer lattice (Figures \ref{fig:int_lat} and \ref{fig:aff_intlat}).  
    This special case has only recently been studied, and it turns out to be very different from the general problem.  
    In particular, there are known polynomial-time algorithms to solve this problem.  
    While not the focus of this paper, a byproduct of our results solves optimal simultaneous flip path problems for lattice triangulations, which we discuss in Section \ref{section:conclusion}.
	
\subsection{Previous Results on Shortest Flip Paths Between Lattice Triangulations}
\label{sec:intro_lat}
	 
    The first $O(n^2)$-time algorithm to find a shortest flip path between two lattice triangulations of an $n$ point-set was given in  \cite{eppstein2007happy}. 
    This algorithm also solves the more general problem of finding a shortest flip path between two (not necessarily lattice) triangulations of a point-set that contains no empty, strictly convex pentagon and whose bounding polygon is convex.  
    Lattice point-sets bounded by convex polygons have this property.  
    
    Another $O(n^2)$-time algorithm for this problem was presented in \cite{caputo2015}.  
    The motivation in this paper was to use Markov chains to generate (and count) random lattice triangulations, which define a spin system.  
    The algorithm does not require that the bounding polygon of the lattice point-set be convex.
    Furthermore, it finds a shortest flip path even when all triangulations along the path must contain a given set of point-pairs - i.e., pairs of points - as edges.  
    These point-pairs are referred to as \emph{constraint edges} and the flip path is called \emph{constrained}.  
    The fact that constrained flip paths always exists between lattice triangulations is not trivial, and is proved in \cite{caputo2015}, along with an upper-bound on the length of a constrained shortest flip path between lattice triangulations whose point-sets are \emph{rectangular}, i.e. a set $S=\{(x,y) \in \mathbb{Z}^2: 0 \leq x \leq l \text{ and } 0 \leq y \leq m\}$, where $l$ and $m$ are any positive integers.
    
    \begin{lemma}[\cite{caputo2015}, Lemma 3.7]
    \label{lem:caputo_bound}
        The length of a constrained shortest flip path between two lattice triangulations of a rectangular lattice point-set with $n$ points is $O(n^\frac{3}{2})$.
    \end{lemma}

    Another important result from  \cite{caputo2015} is that an edge in a lattice triangulation can be flipped to a shorter edge if and only if the bounding point-pairs of its \emph{minimum parallelogram} are edges in the triangulation.  
    We call this parallelogram the \emph{Farey parallelogram for the edge} (Definition \ref{def:farey_parallelogram}), because of its relation to Farey sequences \cite{Farey}, as detailed in Section \ref{sec:farey_plan}.  
    For example, the four edges bounding each red edge in Figures \ref{fig:flip} - \ref{fig:sec1_fp5} all form Farey parallelograms, since the red edges can clearly be flipped to shorter edges.
    
    Both \cite{eppstein2007happy} and \cite{caputo2015} prove the uniqueness of the set of flips in any shortest flip path between two lattice triangulations.  
    However, the structure of this set has not been studied.  
    We give algorithms that compute a special partial order on this set with the property that each linear ordering of the set that is consistent with the partial order - i.e., each flip in the linear ordering comes after its children in the partial order - is a shortest flip path.  
    While out of the scope of this paper, it is not difficult to show that the converse is also true, i.e., every shortest flip path is consistent with this partial order. 
    
    Lastly, although the point-sets of lattice triangulations in \cite{eppstein2007happy} and \cite{caputo2015} are all of the form $S=L\cap\Omega$, where $L$ is the integer lattice $\mathbb{Z}^2$, it is easy to see that the results in these papers apply even when $L$ is any affine transformation of $\mathbb{Z}^2$.  
    Additionally, the algorithm in \cite{eppstein2007happy} can be modified to handle non-convex bounding polygons and constraint edges while achieving the same time complexity.  
    
\section{Contributions and Guide to Reading}
\label{sec:contributions}

    From here on, all triangulations are lattice triangulations of a lattice point-set $S = L \cap \Omega$.  
    Recall that we let $n = |S|$.  
    
    \begin{note}
        Since a lattice point-set is a finite subset of a lattice, much of the geometric, algebraic, and partially ordered set (poset) structure of lattices does not play a role.  
        However, we judiciously exploit a connection between lattices and Farey sequences \cite{Farey}.  
    \end{note}
    
    We study the following problem: given two triangulations of a lattice point-set and a set $F$ of constraints edges (see Section \ref{sec:intro_lat}) in both triangulations, compute the shortest flip path constrained by $F$ between these triangulations.  
    The following main theorems are proved in Section \ref{section:shortest_path}.
    
    \begin{theorem}[Unique Constrained Shortest Flip Paths]
    \label{thm:main_fp}
        The shortest constrained flip path between two lattice triangulations of a lattice point-set is unique, up to reordering flips.
    \end{theorem}
    
    In Section \ref{section:shortest_path}, we give a large class of inputs for which the shortest flip path in the above theorem can be found in time linear in its length (Proposition \ref{prop:output_sensitive_complexity}).  
    We sketch the proof of Theorem \ref{thm:main_fp} below.  
    
    \begin{theorem}[Complexity for Rectangular Lattice Point-sets]
    \label{thm:main_rectangular_complexity}
        There is an $O(n^\frac{3}{2})$-time algorithm to compute the shortest constrained flip path   between two lattice triangulations of an $n$-point rectangular lattice point-set.  
    \end{theorem}
    
    Theorem \ref{thm:main_rectangular_complexity} bridges the gap between the $O(n^2)$-time algorithms in \cite{eppstein2007happy} and \cite{caputo2015} and the $\Omega(n^{\frac{3}{2}})$-time lower-bound implied by Lemma \ref{lem:caputo_bound}.  
    The proof of this theorem follows from Proposition \ref{prop:overall_complexity} (Section \ref{section:shortest_path}) and Lemma \ref{lem:caputo_bound} (Section \ref{section:previous_work}).
    In the next section, we compare our algorithm to those in \cite{eppstein2007happy} and \cite{caputo2015}.  
    
    Next, we sketch the proof ideas of Theorem \ref{thm:main_fp}.  
    First, we show (Proposition \ref{prop:bij_tri_min_2}) that any triangulation $T$ of a lattice point-set $S$ is a so-called \emph{minimum triangulation of $S$} (Definition \ref{def:minimum_tri}), which is a unique triangulation (Lemma \ref{lem:unique_minimum_tri}) containing a  certain minimum subset $G$ of edges in $T$.
    Specifically, in Section \ref{sec:ground_state}, we characterize the minimum set $G$ of edges in $T$ that uniquely determines $T$ (Definition \ref{def:ind_in_S}).  
    We observe an analogy to (the equivalence class of) \emph{ground state triangulations constrained by $G$} defined in \cite{caputo2015} as triangulations of $S$ that contain $G$ as edges and minimize the sum of the $l_1$-lengths of their edges.
    Thus, the theorem is proved by showing that it holds for any two minimum triangulations (Theorem \ref{thm:constr_min_flp}).  
    This step relies on our new structural result for flip paths between lattice triangulations, which is our third main theorem, proved in Section \ref{section:multi_edge}.  
    We require the following definitions to state this theorem.
    
    \begin{figure}[tbhp]
        \centering
        \begin{subfigure}[t]{0.35\linewidth}
            \centering
            \includegraphics[width=\linewidth]{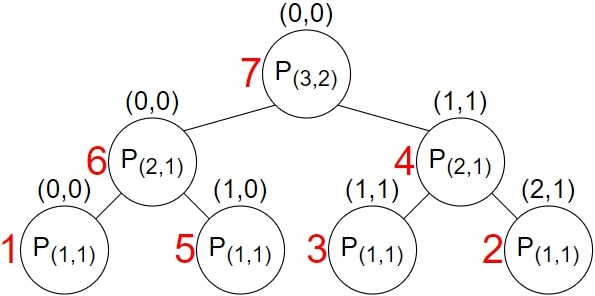}
            \caption{ }
        \label{fig:flip_plan_3_2}
        \end{subfigure}
        \begin{subfigure}[t]{0.15\linewidth}
            \centering
            \includegraphics[width=\linewidth]{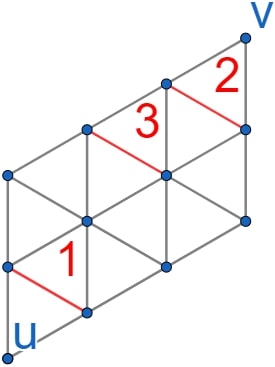}
            \caption{ }
        \label{fig:force_point-pair_1}
        \end{subfigure}
        \begin{subfigure}[t]{0.15\linewidth}
            \centering
            \includegraphics[width=\linewidth]{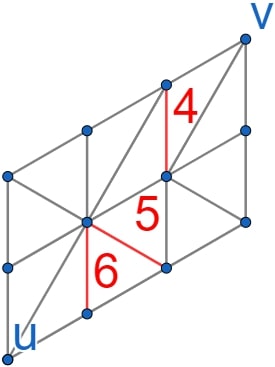}
            \caption{ }
        \label{fig:force_point-pair_2}
        \end{subfigure}
        \begin{subfigure}[t]{0.15\linewidth}
            \centering
            \includegraphics[width=\linewidth]{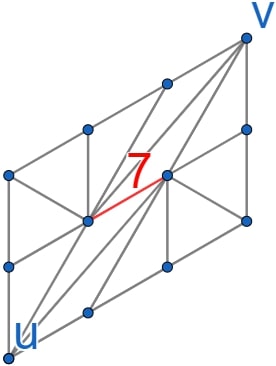}
            \caption{ }
        \label{fig:force_point-pair_3}
        \end{subfigure}
        \begin{subfigure}[t]{0.15\linewidth}
            \centering
            \includegraphics[width=\linewidth]{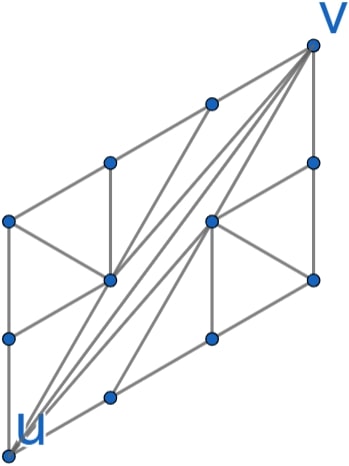}
            \caption{ }
        \label{fig:force_point-pair_4}
        \end{subfigure}
        \caption{(a) A minimum flip plan (Definition \ref{def:flip_plan}) that starts from an equilateral triangulation and forces the point-pair $(u,v)$ in (b) to become an edge.  
        Each node is a flip on a Farey parallelogram (Definition \ref{def:farey_parallelogram}).  
        The red numbers indicate the order of flips in a consistent linear ordering of the flip plan, which is a flip path.  
        (b)-(e) Triangulations along this flip path.  
        The red numbers correspond to those in (a) and indicate the edges removed by each flip in the flip plan.  
        See below and Section \ref{section:single_edge}.}
    \label{fig:flip_plan_cons}
    \end{figure}
    
    A \emph{consistent linear ordering} of a poset, or a \emph{linear ordering consistent with} a poset, is a linear ordering of the elements of the poset such that each element comes after its children.  
    For example, see Figure \ref{fig:flip_plan_cons}.  
    
    \begin{definition}[Flip plan]
    \label{def:flip_plan}
        A \emph{flip plan} is a poset of flips with the property that all of its consistent linear orderings are flip paths between the same two triangulations.  
        The starting and target triangulations of a flip plan are the starting and target triangulations of these flip paths.
    \end{definition}
    
    For example, a flip path is a flip plan with a unique consistent linear ordering.  
    We are interested in \emph{minimum} flip plans, i.e., flip plans whose consistent linear ordering are all shortest flip paths.  
    Figure \ref{fig:flip_plan_cons} shows such a flip plan along with one of its consistent linear orderings, which is a flip path, and triangulations along this path.  
    Note that all the flips in each level of this flip plan can be performed simultaneously.  
    
    The \emph{equilateral triangulation} of a lattice point-set $S$, if it exists, is the triangulation that contains only unit-length edges (Definition \ref{def:unit_edge}).  
    For example, see Figure \ref{fig:force_point-pair_1}.  
    A lattice point-set that admits an equilateral triangulation is an \emph{equilateral lattice point-set}.  
    
    Finally, a flip path that starts from a triangulation $T$ of an equilateral lattice point-set $S$ and \emph{forces a point-pair $(u,v)$ in $S$ to become an edge} is a flip path between $T$ and \emph{some} triangulation in which $(u,v)$ is an edge.  
    This flip path is shortest if it also has minimum length over all target triangulations in which $(u,v)$ as an edge.
    If the consistent linear orderings of a flip plan are all flip paths of this type, then the flip plan starts from $T$ and \emph{forces $(u,v)$ to become an edge}.  
    
    For example, Figure \ref{fig:flip_plan_cons} shows a minimum flip plan that starts from an equilateral triangulation and forces the point-pair $(u,v)$ to become an edge.  
    Also, returning to the shortest flip path between the triangulations in Figures \ref{fig:sec1_fp1} and \ref{fig:sec1_fp5}, the following $3$ sets of flips can be performed simultaneously: 
    those that add the red edges in Figure \ref{fig:sec1_fp3}, those that add the red edges in Figure \ref{fig:sec1_fp4}, 
    and the two that add the $2$ red edges in Figure \ref{fig:sec1_fp5}.  
    These sets hint at a minimum flip plan that starts from the equilateral triangulation in Figure \ref{fig:sec1_fp1} and forces the point-pairs of the last $2$ red edges to become edges.
    
    \begin{theorem}[A Minimum Flip Plan that Starts from an Equilateral Triangulation and Forces a Point-pair to Become an Edge]
    \label{thm:main_flp}
        There is an algorithm that computes a minimum flip plan that starts from an equilateral triangulation and forces a point-pair to become an edge, and it runs in time linear in the number of flips in the flip plan.
    \end{theorem}
    
    The algorithm in Theorem \ref{thm:main_flp} is the fundamental tool that we use to compute minimum flip plans between two triangulations.  
    To the best of our knowledge, this algorithm is the first to exploit the relationship between lattices and Farey sequences \cite{Farey}.
    Most of the work in this paper is focused on generalizing Theorem \ref{thm:main_flp}.  
    In Section \ref{section:multi_edge}, we generalize to a set of point-pairs (Theorem \ref{thm:MultiEdgeCreate}).  
    Then, in Section \ref{sec:ground_state}, we allow the starting triangulation to be a minimum triangulation (Theorem \ref{thm:constrained_flp_not_equilateral}).
    
    Notably, all the above results follow from simple number-theoretic and geometric concepts, such as the connection between lattices and Farey sequences.  
    
\subsection{Comparison with Previous Results}
\label{section:previous_work}

    Consider the algorithms in \cite{eppstein2007happy} and \cite{caputo2015} that compute the constrained shortest flip path between two lattice triangulations.  
    Most of the work in these algorithms is in processing the input pair of triangulations.  
    Note that a triangulation of a lattice point-set $S$ is an assignment of edges to the midpoints of $S$ \cite{de2010triangulations}.  
    The algorithm in \cite{eppstein2007happy} constructs a forest such that each tree corresponds to a midpoint of $S$ and each vertex corresponds to an edge assignment.  
    Constructing this forest takes $O(n^2)$ time, where $n = |S|$ (\cite{eppstein2007happy}, Theorem 4).  
    The algorithm in \cite{caputo2015} computes the set of midpoints of $S$ whose edge assignments differ between the input triangulations, which clearly takes $O(n^2)$ time.  
    
    In contrast, the main work in our algorithm is in computing minimum flip plans that force point-pairs to become edges.  
    As a result, our algorithm runs in $O(n^{\frac{3}{2}})$-time (Theorem \ref{thm:main_rectangular_complexity}).  
    Additionally, as mentioned in Section \ref{sec:contributions}, for a large class of inputs, our algorithm runs in time linear in the number of flips in the output flip plan (Proposition \ref{prop:output_sensitive_complexity}).  

\section{Basic Tools and Definitions}
\label{section:preliminaries}

    In this section, we present the definitions and notation that will be used throughout the paper and we explore an important relationship between lattices and Farey sequences \cite{Farey}.  
    
    \begin{figure}[tbhp]
        \centering
        \includegraphics[width=0.3\linewidth]{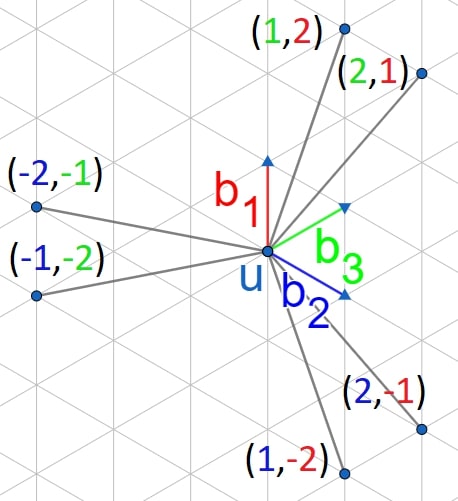}
        \caption{A three-direction lattice and its pseudo-basis $\{b_1,b_2,b_3\}$ along with point-pairs originating at the point $u$ represented using their respective defining coordinate pairs (Definition \ref{def:defining_coordinate_pair}).  
        All the point-pairs belong to the equivalence class $(1,2)$ (Definition \ref{def:equiv}).  
        See below and Section \ref{sec:farey_plan}.}
    \label{fig:edge_3coord}
    \end{figure}
    
    Let $L$ be an affine transformation of the integer lattice $\mathbb{Z}^2$.  
    We can treat $L$ as a three-direction lattice as follows.
    Consider the standard basis vectors $b_1$ and $b_2$ of $L$.  
    Let $b_3$ be the shorter of $b_1-b_2$ and $b_1 + b_2$.  
    If neither vector is shorter, then choose one arbitrarily.  
    The set $\{b_1,b_2,b_3\}$ is the \emph{pseudo-basis} of $L$.  
    Observe that any vector $g$ in $L$ can be written as an integer linear combination of any two distinct pseudo-basis vectors $b_i$ and $b_i$, i.e., $g = xb_i + yb_j$, where $x$ and $y$ are integers.  
    Also, note that we only care about the values of the coefficients $x$ and $y$, and never the values of any pseudo-basis vector.  

    \begin{definition}[Defining Coordinate Pair]
    \label{def:defining_coordinate_pair}
        Let $g$ be a vector in a three-direction lattice $L$ whose pseudo-basis is $\{b_1,b_2,b_3\}$.  
        If $g$ is not a pseudo-basis vector, then its \emph{defining coordinate pair} is the counter-clockwise ordered pair $(b_i,b_j)$ of distinct pseudo-basis vectors such that $g = xb_i + yb_j$ and $|x|+|y|$ is minimized.
        Otherwise, the \emph{defining coordinate pair} of $g$ is any counter-clockwise ordered pair $(b_i,b_j)$ of distinct pseudo-basis vectors such that $g$ is either $b_i$ or $b_j$.  
    \end{definition}

    Note that the defining coordinate pair of a pseudo-basis vector $g$ is not unique, by definition, however one coordinate of any such pair is fixed and carries all the information we need to uniquely identify $g$.  

    \begin{definition}[A Point-pair as a Pair of Vectors]
    \label{def:edge_point}
        We represent a point-pair $(u,v)$ in a three-direction lattice $L$ as $(g,u)$, where $g=v-u$ is a vector represented using its defining coordinate pair $(b_i,b_j)$, i.e., $g$ is written as $(x,y)$ where $g = xb_i + yb_j$.  
        We refer to $g$ as the \emph{vector}, $u$ as the \emph{originating point}, and $(b_i,b_j)$ as the defining coordinate pair of $(g,u)$.  
        Additionally, the \emph{line-segment of} $(g,u)$ is the line-segment between $u$ and $v$.
    \end{definition}

    Consider the point-pairs in Figure \ref{fig:edge_3coord} that originate from the same point $u$.  
    Each of these point-pairs is represented using its defining coordinate pair, which is indicated by the colors of the coordinates.  
    Given a vector $g$, an originating point $u$, and a defining coordinate pair, the point-pair $(g,u)$ is unique.  
    Throughout this paper, the defining coordinate pair of a point-pair is always clear from context and is almost never explicitly specified or used.  
    Similar representations of point-pairs using vectors in three-direction lattices appear in the study of box-splines \cite{benhfid2020reversible,condat2006three}.
    
    \begin{note}
        Unless stated otherwise, all vectors discussed from here on are vectors in a three-direction lattice.  
        If a vector is expressed with only two coordinates, it is assumed to be represented using its defining coordinate pair.  
        Furthermore, if a point-pair is an edge of a lattice triangulation, then the coordinates of its vector either are relatively prime or each has magnitude at most $1$, as discussed below.  
        Thus, unless stated otherwise, the coordinates of any vector discussed either are relatively prime or each has magnitude at most $1$.
    \end{note}
    
    Finally, consider any point-pair $(g,u)$ in a lattice point-set $S$.  
    If the coordinates of $g$ are not relatively prime and the magnitude of at least one coordinate is greater than $1$, then it is easy to see that some point of $S$ other than $u$ or $u + g$ is incident on the line-segment of $(g,u)$.  
    Hence, combining this with the fact that triangulations are embeddings of graphs into $\mathbb{R}^2$ which are \emph{maximally planar} implies that $(g,u)$ is an edge of some triangulation of $S$ if and only if the coordinates of $g$ either are relatively prime or each has magnitude at most $1$.
    In the next section, we explore how this property connects edges in triangulation to fractions in Farey sequences \cite{Farey}.  

\subsection{The Farey Plan for a Vector}
\label{sec:farey_plan}

    In this section, we give an algorithm that maps a vector $g$ to a sequence of fractions obtained from Farey sequences, called the \emph{Farey plan} for $g$.  
    First, we define Farey sequences and state two of their properties.  
    
    \begin{figure}[tbhp]
        \centering
        \includegraphics[width=0.25\linewidth]{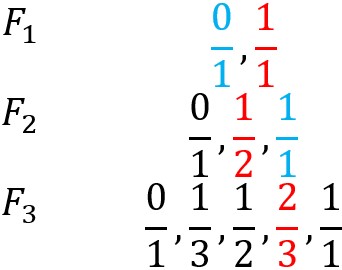}
        \caption{Farey sequences (Definition \ref{def:farey_sequence}) of orders $1$, $2$, and $3$.  
        The pairs of blue and red fractions in each Farey sequence are those in the sequence described in Lemma \ref{lem:farey_intervals} for the fraction $\frac{2}{3}$.  
        The fractions in the Farey plan (Definition \ref{def:farey_plan}) for any vector in the equivalence class $(2,3)$ is shown in red.  
        See below.}
    \label{fig:farey_plan}
    \end{figure}
    
    \begin{definition}[Farey Sequence]
    \label{def:farey_sequence}
        A \emph{Farey sequence \cite{Farey} of order $m$}, denoted by $F_m$, is a strictly-increasing sequence containing $\frac{0}{1}$, $\frac{1}{1}$, and all fully-reduced fractions (relatively prime pairs of nonnegative integers) in the range $[0,1]$ with denominators of at most $m$.  
        Given a fully-reduced fraction $f = \frac{x}{y}$ with $0 \leq f \leq 1$, we let $F_f$ denote the lowest order Farey sequence containing $f$, which is $F_y$.
    \end{definition}
    
    \begin{property}[Mediant of Farey Neighbors \cite{Farey}]
    \label{property:farey_mediant}
        For all $m \geq 1$, each fraction $f_i = \frac{x}{y}$ in the Farey sequence $F_m$, other than $\frac{0}{1}$ and $\frac{1}{1}$, is the mediant of its \emph{Farey neighbors}, which are the fractions $f_{i-1} = \frac{a}{b}$ and $f_{i+1} = \frac{c}{d}$, i.e., $\frac{x}{y} = \frac{a+c}{b+d}$.  
    \end{property}
    
    \begin{property}[Computing Farey Sequences \cite{Farey}]
    \label{property:farey_compute}
        The Farey sequence $F_{i+1}$ contains the fractions in the Farey sequence $F_i$ and the mediant of each pair of consecutive fractions in $F_i$ whose denominators sum to $i+1$.
    \end{property}
    
    For examples, see Figure \ref{fig:farey_plan}.  
    As noted at the end of the previous section, if a point-pair $(g,u)$ is an edge of a triangulation, then the coordinates of $g$ either are relatively prime or each has magnitude at most $1$.  
    We will use this fact to map $g$ to a fraction in a Farey sequence.  
    Then, we will use Properties \ref{property:farey_mediant} and \ref{property:farey_compute} to construct the Farey plan for $g$ (Definition \ref{def:farey_plan}), which will be used in the next section as a roadmap to construct a flip plan.  
    
    Next, we prove several properties about Farey sequences.  
    
    \begin{lemma}[Unique Intervals Containing a Fraction]
    \label{lem:farey_intervals}
        Let $f$ be a fraction and let $m \geq 2$ be an integer such that $F_f=F_m$.  
        There exists a sequence $(f_{1,1},f_{1,2}),\dots,\\(f_{m-1,1},f_{m-1,2})$ of pairs of fractions such that, for each $1 \leq i \leq m-1$, $(f_{i,1},f_{i,2})$ is the unique pair of fractions that are adjacent in $F_i$ with $f_{i,1} < f < f_{i,2}$.  
        Furthermore, if $f'$ is the mediant of $f_{i,1}$ and $f_{i,2}$, then either 
        \begin{enumerate}
            \item $f'=f$ or
            \item the first pair in the above sequence that is not equal to $(f_{i,1},f_{i,2})$ is either $(f_{i,1},f')$ or $(f',f_{i,2})$.  
        \end{enumerate}
    \end{lemma}
    
    \begin{proof}
        Since the fractions in each Farey sequence are ordered from least to greatest, the first part of the lemma is immediate.  
        For the second part, consider the pair $(f_{i,1},f_{i,2})$, for any $1 \leq i \leq m-1$, and let $f'$ be the mediant of this pair.  
        By Property \ref{property:farey_compute}, $\{f_{i,1},f',f_{i,2}\}$ is a contiguous subsequence of $F_{f'}=F_j$, where $j > i$.  
        Furthermore, $f'$ is the only fraction in any Farey sequence of order at most $j$ that lies between $f_{i,1}$ and $f_{i,2}$.  
        Thus, either $f = f'$ or the first pair in the sequence given in the lemma statement that is not equal to $(f_{i,1},f_{i,2})$ is either $(f_{i,1},f')$ or $(f',f_{i,2})$.  
    \end{proof}
    
    \begin{lemma}[Sequence of Mediants]
    \label{lem:farey_plan}
        Consider a fraction $f$ such that the Farey sequence $F_f$ has order at least $2$.  
        Also, consider the sequence of pairs of fractions given by Lemma \ref{lem:farey_intervals} for $f$ and let 
        $$(f_{1,1},f_{1,2}),\dots,(f_{n,1},f_{n,2})$$
        be the subsequence of all distinct pairs.  
        Finally, let $f_0,f_1,\dots,f_n$ be the sequence of fractions where $f_0=\frac{1}{1}$ and $f_i$ is the mediant of $(f_{i,1},f_{i,2})$, for all $1 \leq i \leq n$.  
        Then, $f_n = f$ and $f_i$ is the mediant of $f_{i-1}$ and a Farey neighbor of $f_{i-1}$ in the Farey sequence $F_{f_{i-1}}$, for all $1 \leq i \leq n$.  
    \end{lemma}
    
    \begin{proof}
        By Lemma \ref{lem:farey_intervals}, we have $f_n=f$.  
        Next, consider the fractions $f_{i-1}$ and $f_i$, for any $1 \leq i \leq n$.  
        If $i=1$, then $f_1=\frac{1}{2}$ and the lemma is immediate.  
        Otherwise, Lemma \ref{lem:farey_intervals} tells us that $(f_{i,1},f_{i,2})$ is either $(f_{i-1,1},f_{i-1})$ or $(f_{i-1},f_{i-1,2})$.  
        Furthermore, by Property \ref{property:farey_compute}, $f_{i-1,1}$ and $f_{i-1,2}$ are the Farey neighbors of $f_{i-1}$ in the Farey sequence $F_{f_{i-1}}$.  
        Therefore, $f_i$ is the mediant of $f_{i-1}$ and a fraction adjacent to $f_{i-1}$ in $F_{f_{i-1}}$.  
    \end{proof}
    
    The following is a map between vectors and fractions in Farey sequences, which allows us to define equivalence relations on vectors and point-pairs.
    
    \begin{definition}[Farey-Flip Map]
    \label{def:farey_flip_map}
        Given a vector $g = (x,y)$, the \emph{Farey-Flip map} $\phi_g$ sends $g$ to the fraction $\frac{|x|}{|y|}$ if $|x|\leq |y|$, or to the fraction $\frac{|y|}{|x|}$ otherwise.  
        For any vector $g' = (x',y')$, $\phi_g$ sends $g'$ to the fraction $\frac{|x'|}{|y'|}$ if $\phi_g(g) = \frac{|x|}{|y|}$, or to the fraction $\frac{|y'|}{|x'|}$ otherwise.  
        
        The inverse Farey-Flip map $\phi^{-1}_g$ takes a fraction $\frac{a}{b}$ to a vector $g''=(c,d)$ such that (i) $g''$ and $g$ have the same defining coordinate pair, (ii) the signs on the coordinates of $g''$ and $g$ match, and (iii) $|c|=|a|$ and $|d|=|b|$, if $|x| \leq |y|$, or $|c|=|b|$ and $|d|=|a|$, otherwise.
    \end{definition}
    
    For example, for the vector $g=(-5,3)$, we have $\phi_g(g) = \frac{3}{5}$, $\phi^{-1}_g(\frac{3}{5}) = (-5,3)$, $\phi^{-1}_g \left(\frac{7}{8}\right) = (-8,7)$, and $\phi_g(-8,7) = \frac{7}{8}$, where all vectors have the same defining coordinate pair.  
    The Farey-Flip map allows us to talk about a Farey sequence \emph{containing a vector $g$}, which is a Farey sequence containing the fraction $\phi_g(g)$.  
    We let $F_g$ denote the lowest order Farey sequence containing $g$.  
    In any Farey sequence $F_m$, where $m \geq 1$, we define the \emph{Farey neighbor vectors} of $g$ as the vectors obtained by applying the inverse Farey-Flip map $\phi^{-1}_g$ on the Farey neighbors of $\phi_g(g)$ in $F_m$.  
    
    \begin{definition}[Equivalence Relation on Vectors/Point-pairs]
    \label{def:equiv}
        Two vectors $g$ and $g'$ are equivalent if and only if $\phi_g(g) = \phi_{g'}(g')$, and two point-pairs to be equivalent if and only if their vectors are equivalent.  
        We let $(x,y)$ denote the equivalence class of vectors $g$ (resp. point-pairs $(g,u)$) such that $\phi_g(g) = \frac{x}{y}$.  
    \end{definition}
    
    For example, Figure \ref{fig:edge_3coord} shows point-pairs in the equivalence class $(1,2)$.  
    
    \begin{definition}[Unit-Length Vectors/Point-pairs]
    \label{def:unit_edge}
        A vector (resp. point-pair) is unit-length if it belongs to the equivalence class $(0,1)$.  
    \end{definition}
    
    The above definition formalizes the notion of a unit-length vector, and hence the definition an equilateral triangulation discussed in Section \ref{sec:contributions}.  
    We end this section by defining the Farey plan for a vector and presenting an algorithm to compute it.
    
    \begin{definition}[Farey Plan for a Vector]
    \label{def:farey_plan}
        The \emph{Farey plan $C$ for a vector $g$} is the sequence of fractions defined as follows.  
        If $\phi_g(g) = \frac{0}{1}$, then $C$ is empty.  
        Also, if $\phi_g(g) = \frac{1}{1}$, then $C=\{\frac{1}{1}\}$.  
        Otherwise, $C$ is the sequence for $\phi_g(g)$ given in Lemma \ref{lem:farey_plan}.
    \end{definition}
    
    For example, the Farey plans for the vectors $(-6,1)$ and $(3,5)$ are $\{\frac{1}{1}, \frac{1}{2}, \ldots, \frac{1}{5},\frac{1}{6}\}$ and $\{\frac{1}{1}, \frac{1}{2}, \frac{2}{3}, \frac{3}{5}\}$, respectively.  
    Also, see Figure \ref{fig:farey_plan}.  
    
    \medskip\noindent
    \underbar{Algorithm \emph{Farey Plan}} takes a vector $g = \left(x,y\right)$ as input and outputs the Farey plan $C$ for $g$.  
    The sequence $C$ is initially empty.  
    If $g$ is unit-length, then $C$ is empty.  
    Otherwise, consider the fraction $f=\phi_g(g)$, the sequence $X = F_1 = \left\{\frac{0}{1},\frac{1}{1}\right\}$, and the fraction $d=\frac{1}{1}$.  
    Repeat the following steps until $C$ contains $f$.
    
    \medskip
    \begin{enumerate}
        \item Add the fraction $d$ to the end of $C$.  
        If $C$ contains $f$, then output $C$.
        \item Otherwise, if $f>d$, then assign to $d$ the mediant of $d$ and the fraction directly after $d$ in the sequence $X$; if $f<d$, then assign to $d$ the mediant of $d$ and the fraction directly before $d$ in the sequence $X$.
        \item Insert $d$ into the sequence $X$ such that $X$ is in increasing order.
    \end{enumerate}
    
    \begin{proposition}[Correctness and Complexity of Algorithm Farey Plan]
    \label{prop:complexity_farey_plan}
        On an input vector $g$, Algorithm Farey Plan outputs the Farey plan for $g$ in time linear in the order of the Farey sequence $F_g$.
    \end{proposition}
    
    \begin{proof}
        The proposition is immediate if $g$ belongs to the equivalence class $(0,1)$ or $(1,1)$.  
        Otherwise, observe that Algorithm Farey Plan computes the sequence for $\phi_g(g)$ given in Lemma \ref{lem:farey_plan}.  
        Furthermore, the algorithm clearly runs in time linear in the length of this sequence, which is at most the order of the Farey sequence $F_g$.
    \end{proof}
    
\section{A Minimum Flip Plan for a Point-pair}
\label{section:single_edge}
In this section, we present an algorithm that takes a point-pair and the Farey plan for its vector as inputs and outputs a flip plan that starts from an equilateral triangulation and forces the point-pair to become an edge (until Section \ref{sec:ground_state}, the starting triangulation is always an equilateral triangulation).  
We refer to this as a \emph{flip plan for the point-pair}.  
We start by presenting a convenient representation of flips and flip plans.  

\begin{figure}[tbhp]
    \centering
    \begin{subfigure}[t]{0.45\linewidth}
        \centering
        \includegraphics[width=0.5\linewidth]{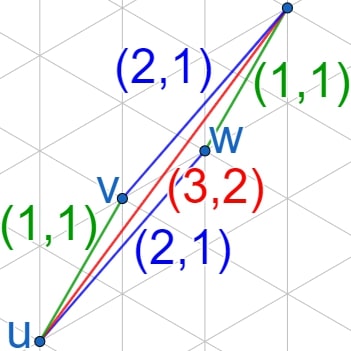}
        \caption{ }
        \label{fig:farey_para_1}
    \end{subfigure}
    \begin{subfigure}[t]{0.45\linewidth}
        \centering
        \includegraphics[width=0.5\linewidth]{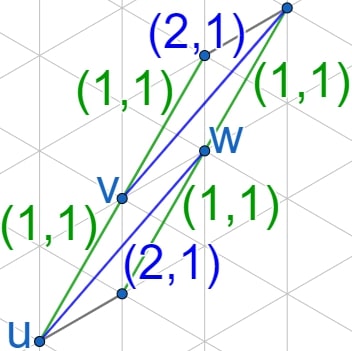}
        \caption{ }
        \label{fig:farey_para_2}
    \end{subfigure}
    \caption{(a) The Farey parallelogram (Definition \ref{def:farey_parallelogram}) for the point-pair $((3,2),u)$, which contains the point-pairs $((2,1),u)$, $((2,1),v)$, $((1,1),u)$, and $((1,1),w)$.  
    (b) The adjacent Farey parallelograms for the point-pairs $((2,1),u)$ and $((2,1),v)$.  
    See below.}
    \label{fig:farey_para}
\end{figure}

\begin{definition}[A Quadrilateral for a Vector/Point-pair]
    A \emph{quadrilateral $Q_g$ for a vector $g$} is a sequence of four vectors $g_1$, $g_2$, $g_3$, and $g_4$ such that $g=g_1+g_2=g_3+g_4$.  
    Notice that the vectors $g$ and $g_1-g_3$ are the diagonals of $Q_g$.
    If $g_1=g_4$ and $g_2=g_3$, then $Q_g$ is called a \emph{parallelogram} and is denoted by $P_g=\{g_1,g_2\}$.  
    
    A \emph{quadrilateral $Q_{g,u}$ for a point-pair} $(g,u)$ is a set of four point-pairs $(g_1,u)$, $(g_2,u+g_1)$, $(g_3,u)$, and $(g_4,u+g_3)$ such that the sequence $g_1,g_2,g_3,g_4$ is a quadrilateral $Q_g$.  
    Since we can obtain $Q_{g,u}$ given the originating point $u$ and $Q_g$, we sometimes refer to $Q_{g,u}$ as $Q_g$ \emph{originating at $u$}.  
    A parallelogram $P_{g,u}$ is defined similarly, and the \emph{longer point-pairs} in $P_{g,u}$ are the two with the longest vectors.
\end{definition}

        In the above definition, note that a quadrilateral $Q_{g,u}$ for a point-pair $(g,u)$ is a quadrilateral whose longer diagonal is $(g,u)$.  
        Hence, any quadrilateral for the shorter diagonal of $Q_{g,u}$ is distinct from $Q_{g,u}$.  
        We define a flip on $Q_{g,u}$ to be the flip that replaces the shorter diagonal of $Q_{g,u}$ with $(g,u)$.  
        We will construct flip plans containing flips on quadrilaterals originating at points in a lattice point-set.  
        More precisely, these flip plans will contain flips on parallelograms of the following type.  
	
	\begin{definition}[The Farey Parallelogram for a Vector/Point-pair]
	\label{def:farey_parallelogram}
	    The \emph{Farey parallelogram} $P_g$ for a non-unit-length vector $g$ is the unique parallelogram for $g$ containing  
	    \begin{enumerate}
	        \item unit-length vectors if $g$ belongs to the equivalence class $(1,1)$, or
	        \item the Farey neighbor vectors of $g$ in the Farey sequence $F_g$ otherwise.
	    \end{enumerate}
	    The Farey parallelogram for a point-pair $(g,u)$ is $P_g$ at $u$.
	\end{definition}
	
		\begin{figure}[htb]
		\centering
		\begin{subfigure}[t]{0.4\linewidth}
			\centering
			\includegraphics[width=\linewidth]{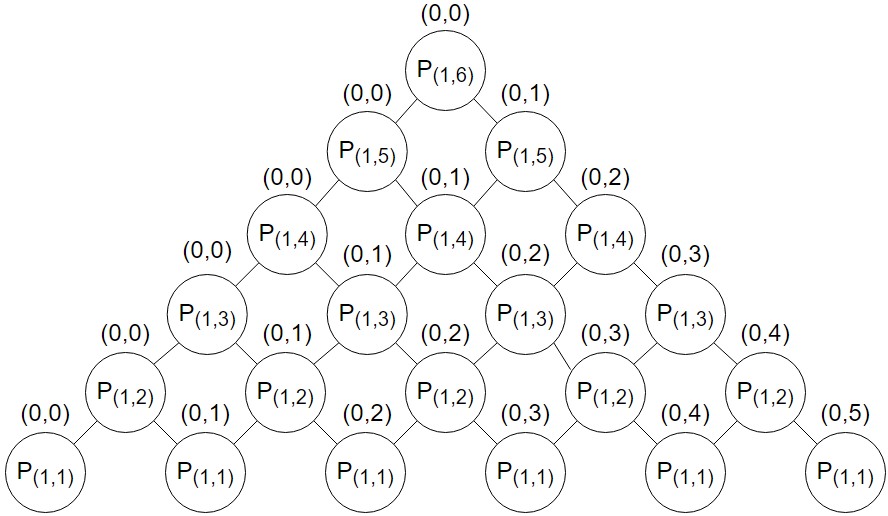}
			\caption{ }
			\label{fig:combining_ops}
		\end{subfigure}
		\begin{subfigure}[t]{0.5\linewidth}
			\centering
			\includegraphics[width=\linewidth]{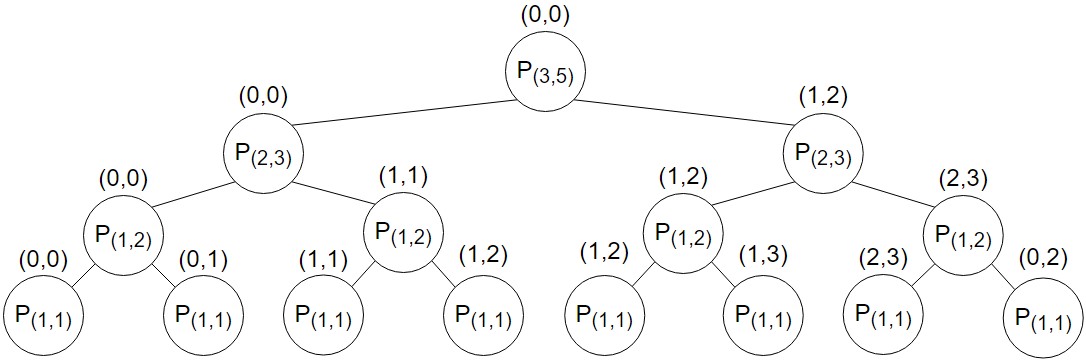}
			\caption{ }
		\end{subfigure}
		\caption{Minimum flip plans for the point-pairs (a) $((1,6),(0,0))$ and (b) $((3,5),(0,0))$.  
		All flips are performed on Farey parallelograms (Definition \ref{def:farey_parallelogram}).  
		Originating points are displayed above each flip.  
		See the discussion below.}
        \label{fig:more_flip_plans}
	\end{figure}

    For example, see Figure \ref{fig:farey_para}.  
 
    We denote a flip plan for a point-pair $(g,u)$ by $\pi_{g,u}(T)$, where $T$ is the starting triangulation.  
    When $T$ is clear from context, we just write $\pi_{g,u}$.  
    The flip that is maximal with respect to the partial order on $\pi_{g,u}$ is called the \emph{maximal flip} of $\pi_{g,u}$.  
    The number of flips in $\pi_{g,u}$ is denoted by $|\pi_{g,u}|$.
	
\begin{figure}[htb]
		\centering
		\begin{subfigure}[t]{0.19\linewidth}
			\centering
			\includegraphics[width=0.8\linewidth]{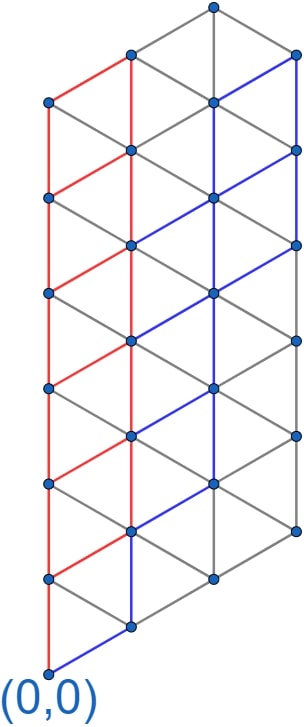}
			\caption{ }
			\label{fig:fp_1}
		\end{subfigure}
		\begin{subfigure}[t]{0.19\linewidth}
			\centering
			\includegraphics[width=0.8\linewidth]{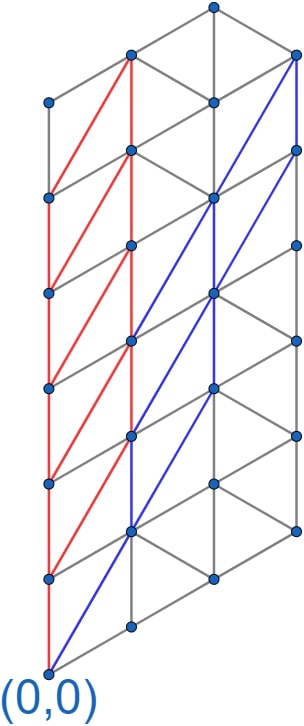}
			\caption{ }
			\label{fig:fp_2}
		\end{subfigure}
		\begin{subfigure}[t]{0.19\linewidth}
			\centering
			\includegraphics[width=0.8\linewidth]{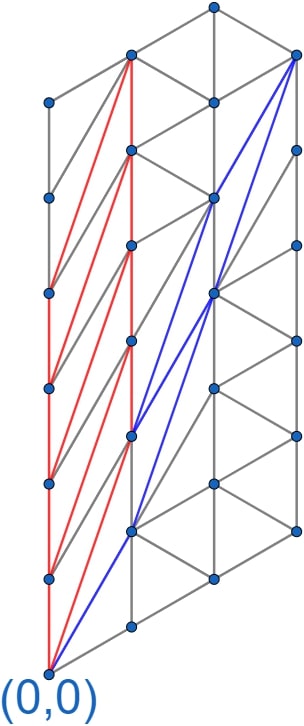}
			\caption{ }
			\label{fig:fp_3}
		\end{subfigure}
		\begin{subfigure}[t]{0.19\linewidth}
			\centering
			\includegraphics[width=0.8\linewidth]{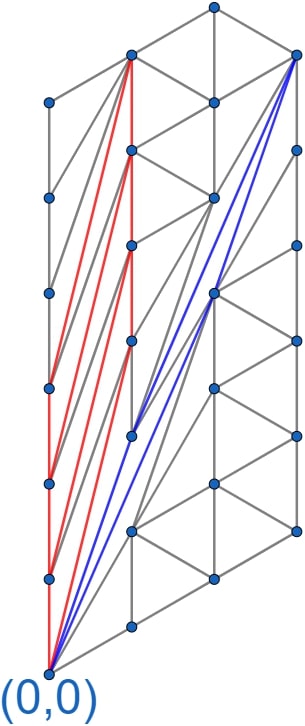}
			\caption{ }
			\label{fig:fp_4}
		\end{subfigure}
		\begin{subfigure}[t]{0.19\linewidth}
			\centering
			\includegraphics[width=0.8\linewidth]{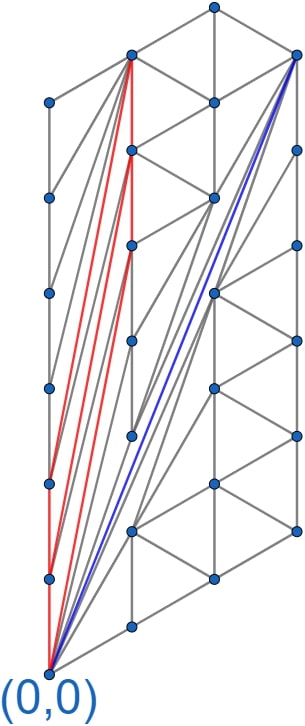}
			\caption{ }
			\label{fig:fp_5}
		\end{subfigure}
		\begin{subfigure}[t]{0.19\linewidth}
			\centering
			\includegraphics[width=0.8\linewidth]{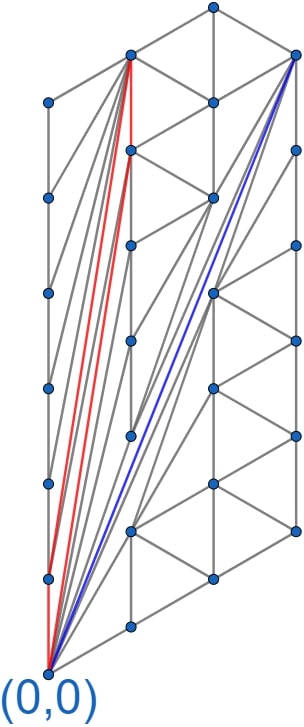}
			\caption{ }
			\label{fig:fp_6}
		\end{subfigure}
		\begin{subfigure}[t]{0.19\linewidth}
			\centering
			\includegraphics[width=0.8\linewidth]{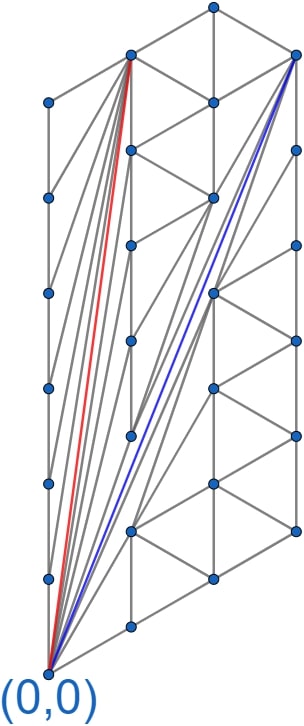}
			\caption{ }
			\label{fig:fp_7}
		\end{subfigure}
		\caption{Triangulations resulting from performing the flips in the flip plans in Figure \ref{fig:more_flip_plans} one level at a time.  
        Flips on Farey parallelograms with at least three red edges are contained in the flip plan for the edge $(1,6)$ at the point $(0,0)$, flips on Farey parallelograms with at least three blue edges are contained in the flip plan for the edge $(3,5)$ at the point $(0,0)$, and flips on Farey parallelograms with two blue and two red edges are contained in both flip plans.}
        \label{fig:farey_parallelograms}
	\end{figure}

    For example, Figure \ref{fig:flip_plan_3_2} shows a minimum flip plan $\pi_{(3,2),(0,0)}$ and Figure \ref{fig:more_flip_plans} shows minimum flip plans (a) $\pi_{(1,6),(0,0)}$ and (b) $\pi_{(3,5),(0,0)}$.  
    These flip plans consist of flips on Farey parallelograms.  
    Also, consider the equilateral triangulation in Figure \ref{fig:fp_1}.  
    Figures \ref{fig:farey_parallelograms} (a)-(f) show all the Farey parallelograms in the latter two flip plans.  
    Figure \ref{fig:fp_7} shows a triangulation containing $((1,6),(0,0))$ and $((3,5),(0,0))$ as edges. 
	
    Next, we prove several properties of Farey parallelograms, stated as Lemmas \ref{lem:fp_neighbor}, \ref{lem:farey_plan_longer_vector}, and \ref{lem:two_halves}, below.  
    Figures \ref{fig:farey_plan} and \ref{fig:farey_para} illustrate these properties.  
	
    \begin{lemma}[Obtaining Farey Parallelograms from Farey Plans]
        \label{lem:fp_neighbor}
	Let $g$ be a vector whose Farey plan is $C = \left\{f_1,\dots,f_n\right\}$, with $n \geq 2$, and let $g_1=\phi_g^{-1}(f_{n-1})$ and $g_2=g-g_1$.  
	Then,
	\begin{enumerate}
	    \item the Farey parallelogram for $g$ is $P_g = \{g_1,g_2\}$ and
	    \item the Farey parallelogram for $g_1$ is $P_{g_1}=\{g_2,g_1-g_2\}$, and hence $g_1$ is longer than $g_2$.
	\end{enumerate}
	\end{lemma}
	
	\begin{proof}
        Using Lemma \ref{lem:farey_plan}, it is easy to see that $g_1$ and $g_2$ are the Farey neighbor vectors of $g$ in $F_g$.  
        Hence, Statement (1) is true.  
        Next, by the same lemma, $g_1$ and $g_2$ are adjacent in $F_{g_1}$.  
        Thus, the Farey neighbor vectors of $g_1$ in $F_{g_1}$ are $g_2$ and $g_1-g_2$, by Property \ref{property:farey_mediant}.  
        This proves Statement (2).
	\end{proof}
	
	\begin{lemma}[Farey Plan for Longer Farey Neighbor Vector]
	\label{lem:farey_plan_longer_vector}
	  Let $g$ be a vector whose Farey plan is $C = \left\{f_1,\dots,f_n\right\}$, with $n \geq 2$, and let $g_1$ be the longer vector in the Farey parallelogram for $g$.  
	  Then, the Farey Plan for $g_1$ is $C \setminus \{f_n\}$.
	\end{lemma}
	
	\begin{proof}
	By Lemma \ref{lem:fp_neighbor}, we have $g_1 = \phi^{-1}(f_{n-1})$.  
	The lemma now follows from the observation that all but the last step of Algorithm Farey Plan on inputs $g$ and $g_1$ are identical.  
 \end{proof}

    To state the next lemma, we require the following terminology.  
    We call two distinct Farey parallelograms $P_{g,u}$ and $P_{g,v}$ \emph{adjacent} if they share a point-pair.  
    We call two distinct point-pairs \emph{adjacent} if their Farey parallelograms are adjacent.
	
	  \begin{lemma}[Decomposing a Farey Parallelogram]
    \label{lem:two_halves}
    Let $(g,u)$ be a point-pair such that the Farey plan for $g$ has length at least $2$, let the longer vector in the Farey parallelogram $P_g$ be $g_1$, and let $g_2=g-g_1$.  
    Also, let $v=u+g_2$ and $w=u+g_1$.  
    Then, the Farey parallelogram $P_{g,u}$ contains the point-pairs $(g_1,u)$, $(g_2,u)$, $(g_1,v)$, and $(g_2,w)$.  
    Furthermore, $P_{g,u}$ is contained in the region bounded by the adjacent Farey parallelograms $P_{g_1,u}$ and $P_{g_1,v}$, $P_{g_1,u}$ contains $(g_2,u)$ and $(g_1-g_2,v)$, and $P_{g_1,v}$ contains $(g_2,w)$ and $(g_1-g_2,v)$.
  \end{lemma}
  
  \begin{proof}
    Using Lemma \ref{lem:fp_neighbor}, we see that the Farey parallelogram $P_{g,u}$ contains the point-pairs $(g_1,u)$, $(g_2,u)$, $(g_1,v)$, and $(g_2,w)$.  
    Next, since the Farey plan for $g$ has length at least $2$, $g$ does not belong to the equivalence classes $(0,1)$ or $(1,1)$.  
    This implies that $g_1$ is not unit-length, and so the Farey parallelogram $P_{g_1}$ exists.  
    By the lemma above, the Farey parallelogram $P_{g_1,u}$ contains $(g_2,u)$.  
    Hence, $P_{g_1,u}$ also contains the point-pair $(g_1-g_2,v)$.  
    Similarly, $P_{g_1,v}$ contains $(g_1-g_2,v)$ and $(g_2,w)$.  
    
    Finally, the discussion above shows that $P_{g_1,u}$ and $P_{g_1,v}$ are adjacent and the region bounded by them contains $P_{g,u}$.  
  \end{proof}

  We are now ready to present our algorithm.  
  
  	\medskip\noindent
	\underbar{Algorithm \emph{Flip Plan}} takes a point-pair $(g,u)$ and the Farey plan $C=\left\{f_1,\dots,f_n\right\}$ for $g$ as inputs and outputs a flip plan $\pi_{g,u}$ for $(g,u)$.  
	The algorithm is recursive and uses a perfect hash table $X$ to keep track of the poset computed thus far to avoid redundant recursive calls.  
 The hash table is initially empty.  

\medskip\noindent
  \emph{Base Cases:} If $C$ is empty, then the poset $\pi_{g,u}$ is empty.  
  If $C = \left\{\frac{1}{1}\right\}$, then add a flip on the Farey parallelogram for the vector $\phi_g^{-1}(\frac{1}{1})$ originating at $u$ to $\pi_{g,u}$.

\medskip\noindent
  \emph{Recursive Step:} Insert $(g,u)$ into $X$ indexed by its vector, defining coordinate pair, and originating point.  
  Consider the vectors $g_1 = \phi_g^{-1}(f_{n-1})$ and $g_2=g-g_1$.  
  Add a flip $s$ on the Farey parallelogram $P_{g} = \left\{g_1,g_2\right\}$ originating at $u$ to $\pi_{g,u}$.  
  The children of $s$ are the maximal flips of the posets that result from recursing on the point-pair $(g_1,u)$ with Farey plan $C \setminus f_n$ and the point-pair $(g_1,u+g_2)$ with Farey plan $C \setminus f_n$.  
  If $X$ contains either $(g_1,u)$ or $(g_1,u+g_2)$, then this recursive call has already been made, and so $s$ is added as a parent to the corresponding maximal flip.
  \medskip

    Given a point-pair, we can compute the Farey plan for its vector.  
    Hence, we often refer to the poset output by Algorithm Flip Plan \emph{on the input point-pair}, omitting mention of the input Farey plan.  
    The remainder of this section is focused on proving the correctness and complexity of this algorithm, which follow from Theorem \ref{thm:construct_two}, stated below.  
    We require the following definitions to state the theorem.
    
    \begin{definition}[Poset Union]
	\label{def:poset_union}
	The \emph{poset union} $\pi \cup \pi'$ of two posets of flips $\pi$ and $\pi'$ is the poset containing all the flips in $\pi$ and $\pi'$ such that two flips are related if and only if they are related in either $\pi$ or $\pi'$.
	\end{definition}
	
	For example, the flip plan that starts from the triangulation in Figure \ref{fig:fp_1} and forces the point-pairs of the blue and red edges in Figure \ref{fig:fp_7} to become edges is the union of the flip plans in Figure \ref{fig:more_flip_plans}.  
	In other words, a single flip plan forces multiple point-pairs to become edges.
	
    Next, two flip plans $\pi_{g,u}$ and $\pi_{g,v}$ are \emph{adjacent} if their maximal flips are on adjacent Farey parallelograms.  
   For example, the two posets in Figure \ref{fig:combining_ops} rooted at the flips on the Farey parallelograms $P_{(1,5),(0,0)}$ and $P_{(1,5),(0,1)}$ are adjacent, as shown in Figure \ref{fig:fp_5}.
	
    \begin{theorem}[A Flip Plan for Adjacent Point-pairs]
    \label{thm:construct_two}
    Let $(g,u)$ and $(g,v)$ be adjacent point-pairs that are edges in some triangulation of an equilateral lattice point-set and let $\pi_{g,u}$ and $\pi_{g,v}$ be the posets output by Algorithm Flip Plan when given these point-pairs as inputs, respectively.  
    Then, $\pi_{g,u} \cup \pi_{g,v}$ is a flip plan for $(g,u)$ and $(g,v)$.
  \end{theorem}

  Theorem \ref{thm:construct_two} is proved at the end of this section using Lemmas \ref{lem:flp_neighbor} and \ref{lem:extend_from_pairs} and a corollary of Lemma \ref{lem:bounding_parallelogram}, below.  
  The following is a corollary of this theorem.  

\begin{corollary}[A Flip Plan for a Point-pair]
  \label{cor:EdgeCreate}
   Let $(g,u)$ be a point-pair that is an edge in some triangulation of an equilateral lattice point-set.  
    Then, on input $(g,u)$, Algorithm Flip Plan outputs a flip plan $\pi_{g,u}$ for $(g,u)$ in $O(|\pi_{g,u}|)$ time.
  \end{corollary}
  
  \begin{proof}
    If $(g,u)$ is unit-length, then it is contained in the equilateral triangulation and $\pi_{g,u}$ is empty, and so the Corollary is true.  
    Otherwise, by Theorem \ref{thm:construct_two}, $\pi_{g,u}$ is a flip plan for $(g,u)$.  
    Lastly, for the complexity statement, we show that the algorithm makes $O(|\pi_{g,u}|)$ recursive calls, each taking $O(1)$ time to complete.  
    This proves the corollary.  
    
    Observe that two point-pairs have the same index in the hash table $X$ if and only if they are the same point-pair.  
    Hence, $X$ is a perfect hash table, and so look-up takes $O(1)$ time.  
    Therefore, it is easy to see that each recursive call takes $O(1)$ time.  
    Finally, each recursive call adds one flip to $\pi_{g,u}$, and so exactly $|\pi_{g,u}|$ recursive calls are made.  
  \end{proof}

    \begin{lemma}[The Structure of a Poset Output by Algorithm Flip Plan]
        \label{lem:flp_neighbor}
        Consider a point-pair $(g,u)$ such that the Farey plan for $g$ has length at least $2$.  
        Let the longer point-pairs in the Farey parallelogram for $(g,u)$ be $(g_1,u_1)$ and $(g_1,u_2)$.  
        Also, let $\pi_{g,u}$, $\pi_{g_1,u_1}$, and $\pi_{g_1,u_2}$ be the posets output by Algorithm Flip Plan when these point-pairs are given as inputs, respectively.  
        Then, $\pi_{g_1,u_1}$ and $\pi_{g_1,u_2}$ are adjacent and $\pi_{g_1,u_1} \cup \pi_{g_1,u_2}$ is the poset $\pi_{g,u}$ with its maximal flip removed.  
    \end{lemma}
    
    \begin{proof}
      Since the Farey plan for $g$ has length at least $2$, the posets $\pi_{g,u}$, $\pi_{g_1,u_1}$, and $\pi_{g_1,u_2}$ are non-empty.  
      Additionally, by Lemma \ref{lem:two_halves}, we have $u_2 = u + g - g_1$.  
      Hence, the two child flips of the maximal flip in $\pi_{g,u}$ are flips on the Farey parallelograms for $(g_1,u_1)$ and $(g_1,u_2)$.  
      Furthermore, by the same lemma, these parallelograms are adjacent.  
      The lemma now follows from the recursive step in Algorithm Flip Plan and Lemma \ref{lem:farey_plan_longer_vector}.
    \end{proof}
    
      \begin{lemma}[Extending Flip Plans]
      \label{lem:extend_from_pairs}
      Let $(g,u)$ and $(g,v)$ be adjacent point-pairs in an equilateral lattice point-set such that the Farey plan for $g$ has length at least $2$.  
      Also, let $(g_1,u_1)$ and $(g_1,u_2)$ be the longer point-pairs in the Farey parallelogram for $(g,u)$ and let $(g_1,u_3)$ and $(g_1,u_4)$ be the longer point-pairs in the Farey parallelogram for $(g,v)$.
      Lastly, on inputs $(g,u)$, $(g,v)$, and $(g_1,u_i)$, let $\pi_{g,u}$, $\pi_{g,v}$, and $\pi_i$ be the posets output by Algorithm Flip Plan, respectively, for all $1 \leq i \leq 4$.
      Then, Statements (1) and (2) imply Statement (3) below:
      \begin{enumerate}
          \item $\pi_i$ is a flip plan for $(g_1,u_i)$, for all $1 \leq i \leq 4$.
          \item $\pi_i \cup \pi_j$ is a flip plan for $(g_1,u_i)$ and $(g_1,u_j)$, for all $1 \leq i,j \leq 4$ with $i \neq j$.
          \item $\pi_{g,u} \cup \pi_{g,v}$ is a flip plan for $(g,u)$ and $(g,v)$.
      \end{enumerate}
    \end{lemma}
    
    The proof of Lemma \ref{lem:extend_from_pairs} can be found in \ref{sec:proof_extend_from_pairs}.  
    
    To state Lemma \ref{lem:bounding_parallelogram}, we need the following definition.  
    Let $(g,u)$ be a point-pair with $g=(x,y)$ that is not unit-length.  
    The \emph{bounding region} for $(g,u)$ is the rectangular lattice point-set whose extreme points are $u$, $u+(x,0)$, $u+(0,y)$, and $u+g$.
	
	\begin{lemma}[Bounding Regions Contain All Flips]
  \label{lem:bounding_parallelogram}
  If a non-unit-length point-pair $(g,u)$ is given as input to Algorithm Flip Plan, then the Farey parallelograms of the flips in the output poset are contained in the bounding region for $(g,u)$.
  \end{lemma}
  
  The proof of Lemma \ref{lem:bounding_parallelogram} can be found in \ref{sec:proof_bounding_parallelogram}.

  \begin{corollary}[A Flip Plan for Point-pairs with Non-intersecting Bounding Regions]
  \label{cor:geom_sep}
  Let $G=\left\{(g_1,u_1),\dots,(g_n,u_n)\right\}$ be a set of point-pairs such that are edges in some triangulation of an equilateral lattice point-set.  
  Also, let the poset $\pi_{g_i,u_i}$ output by Algorithm Flip Plan on input $(g_i,u_i)$ be a flip plan for $(g_i,u_i)$, for all $(g_i,u_i)$ in $G$.  
  If the bounding regions for any two point-pairs in $G$ do not have intersecting interiors, then $\pi_{g_1,u_1} \cup \dots \cup \pi_{g_n,u_n}$ is a flip plan for $G$.
  \end{corollary}

  \begin{proof}
    The corollary follows immediately from Lemma \ref{lem:bounding_parallelogram}.
  \end{proof}
  
\begin{proof}[Proof of Theorem \ref{thm:construct_two}]
    We proceed by induction on the size $n$ of the Farey plan for the vector $g$.  
    Since $(g,u)$ and $(g,v)$ are adjacent, their Farey parallelograms exist, so $n$ is at least $1$.  
    When $n=1$, $(g,u)$ and $(g,v)$ belong to the equivalence class $(1,1)$ and their Farey parallelograms contain unit-length point-pairs.  
    The point-pairs in these Farey parallelograms and the unit-length diagonals of these parallelograms are edges in the equilateral triangulation, and the regions bounded by these parallelograms are disjoint.  
    Hence, flips on both parallelograms can be performed, as specified by $\pi_{g,u} \cup \pi_{g,v}$, and so the base case holds.
    
    Assume that the theorem holds when $n = k$, for any $k \geq 1$, and we will prove it holds when $n=k+1$.  
    Let $(g_1,u_1)$ and $(g_1,u_2)$ be the longer point-pairs in the Farey parallelogram $P_{g,u}$ and let $(g_1,u_3)$ and $(g_1,u_4)$ be the longer point-pairs in the Farey parallelogram $P_{g,v}$, where $u_1=u$ and $u_3=v$.  
    Also, let $\pi_i$ be the posets output by Algorithm Flip Plan on input $(g_1,u_i)$, for all $1 \leq i \leq 4$.  
    It suffices to show that the posets $\pi_i$ satisfy Statements (1) and (2) of Lemma \ref{lem:extend_from_pairs}.  
    Then, applying Lemma \ref{lem:extend_from_pairs} proves the theorem by induction.
    
    To show that Statement (1) is true, consider the poset $\pi_1 \cup \pi_2$.  
    By Lemma \ref{lem:farey_plan_longer_vector}, the Farey plan for the vector $g_1$ has length $k$.  
    Also, by Lemma \ref{lem:flp_neighbor}, $\pi_1 \cup \pi_2$ is $\pi_{g,u}$ with its maximal flip removed, and the maximal flips in $\pi_1 \cup \pi_2$ are on the adjacent Farey parallelograms for $(g_1,u_1)$ and $(g_1,u_2)$.  
    Hence, by the inductive hypothesis, $\pi_1 \cup \pi_2$ is a flip plan for $(g_1,u_1)$ and $(g_1,u_2)$.  
    Similarly, $\pi_3 \cup \pi_4$ is a flip plan for $(g_1,u_3)$ and $(g_1,u_4)$.
    Therefore, $\pi_i$ is a flip plan for $(g_1,u_i)$, for all $1 \leq i \leq 4$.

\begin{figure}
    \centering
    \begin{subfigure}[t]{0.13\linewidth}
        \centering
        \includegraphics[width=\linewidth]{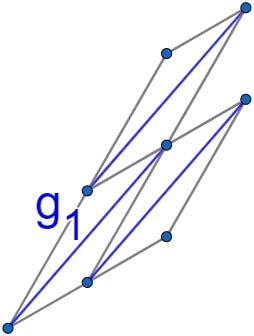}
        \caption{ }
    \end{subfigure}
    \begin{subfigure}[t]{0.13\linewidth}
        \centering
        \includegraphics[width=\linewidth]{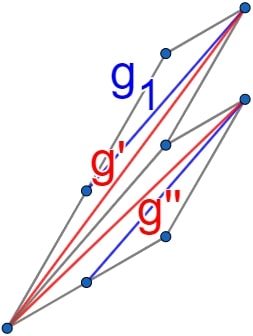}
        \caption{ }
    \end{subfigure}
    \begin{subfigure}[t]{0.17\linewidth}
        \centering
        \includegraphics[width=1.1\linewidth]{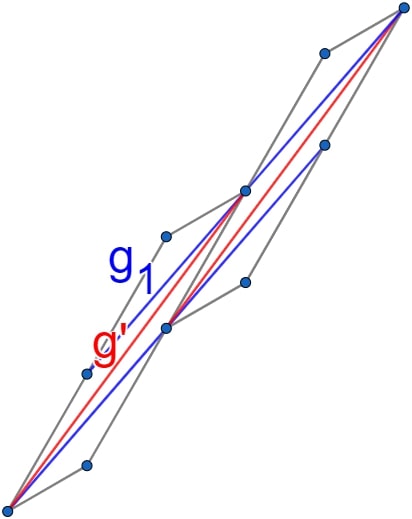}
        \caption{ }
    \end{subfigure}
    \begin{subfigure}[t]{0.17\linewidth}
        \centering
        \includegraphics[width=0.9\linewidth]{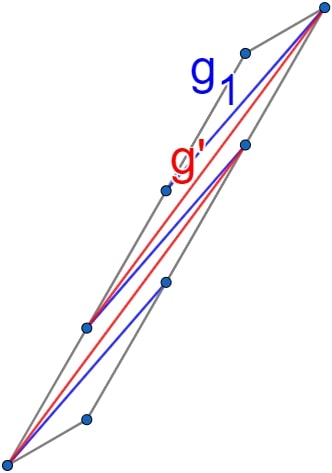}
        \caption{ }
    \end{subfigure}
    \begin{subfigure}[t]{0.16\linewidth}
        \centering
        \includegraphics[width=1.1\linewidth]{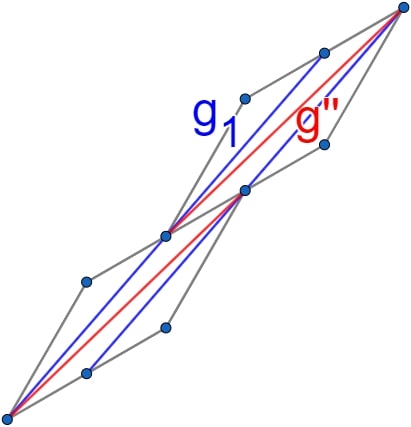}
        \caption{ }
    \end{subfigure}
    \begin{subfigure}[t]{0.16\linewidth}
        \centering
        \includegraphics[width=\linewidth]{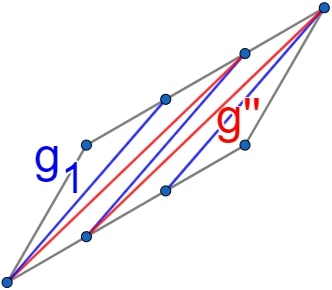}
        \caption{ }
    \end{subfigure}
    \caption{(a) The two cases for the equivalence class of the point-pair shared between adjacent Farey parallelograms whose longer diagonals belong to the equivalence class $g_1$.  
    (b) The cases $g'$ and $g''$ for the equivalence class of the vector $g$ such that the longer diagonal of the Farey parallelogram for $g$ is $g_1$.  
    See the proof of Theorem \ref{thm:construct_two}.  
    (c)-(f) The four cases for the configuration of adjacent point-pairs such that their vectors are $g$ and the longer vector in the Farey parallelogram for $g$ is $g_1$.  
    These cases result from the cases in (a) and (b).  
    See the proof of Theorem \ref{thm:construct_two}.}
    \label{fig:stronger_claim}
\end{figure}
    
    Next, we show that Statement (2) is true.  
    We have already shown that $\pi_1 \cup \pi_2$ is a flip plan for $(g_1,u_1)$ and $(g_1,u_2)$ and $\pi_3 \cup \pi_4$ is a flip plan for $(g_1,u_3)$ and $(g_1,u_4)$.  
    For the other posets, consider the adjacent Farey parallelograms $P_{g_1,u_1}$ and $P_{g_1,u_2}$.  
    Note that there are two cases for the equivalence class of the point-pair shared between $P_{g_1,u_1}$ and $P_{g_1,u_2}$.  
    These cases correspond to two cases for the vector $g_1$, as shown in Figure \ref{fig:stronger_claim} (a) and (b).  
    Likewise, there are two cases for the equivalence class of the point-pair shared between $P_{g,u}$ and $P_{g,v}$.  
    Therefore, without loss of generality, there are four cases for the configuration of $(g,u)$ and $(g,v)$ in the lattice point-set, as shown in Figure \ref{fig:stronger_claim}.  
    
    In cases (c) and (e), the shared point-pair belongs to the equivalence class $g-g_1$.  
    Hence, the point-pairs $(g_1,u_1)$, $(g_1,u_2)$, $(g_1,u_3)$, and $(g_1,u_4)$ are distinct and, without loss of generality, the point-pairs $(g_1,u_2)$ and $(g_1,u_3)$ are adjacent.  
    By the inductive hypothesis, $\pi_2 \cup \pi_3$ is a flip plan for $(g_1,u_2)$ and $(g_1,u_3)$.  
    Also, observe that $(g_1,u_3)$ and $(g_1,u_4)$ can be obtained from $(g_1,u_1)$ by translating its origin point by $g_1$ and $g$, respectively.  
    Similarly, $(g_1,u_4)$ can be obtained from $(g_1,u_2)$ by translating its origin point by $g_1$.  
    Since bounding regions are convex by definition, this implies that the bounding regions for the point-pairs in each pair $((g_1,u_1),(g_1,u_3))$, $((g_1,u_1),(g_1,u_4))$, and $((g_1,u_2),(g_1,u_4))$ do not have intersecting interiors.  
    Therefore, by Corollary \ref{cor:geom_sep}, $\pi_1 \cup \pi_3$, $\pi_1 \cup \pi_4$, and $\pi_2 \cup \pi_4$ are flip plans for their respective pairs of point-pairs.  
    
    Finally, in cases (d) and (f), the shared point-pair belongs to the equivalence class $g_1$.  
    Hence, without loss of generality, the point-pairs $(g_1,u_2)$ and $(g_1,u_3)$ are the same.  
    Hence, by the argument above, $\pi_1 \cup \pi_2$ and $\pi_2 \cup \pi_3$ are flip plans are flip plans for their respective pairs of point-pairs.
    Lastly, by transitivity, the poset $\pi_1 \cup \pi_3$ is a flip plan $(g_1,u_1)$ and $(g_1,u_3)$.  
    Thus, Statement (2) of Lemma \ref{lem:extend_from_pairs} is true, and the proof is complete.  
  \end{proof}

\section{A Minimum Flip Plan for a Set of Point-pairs}
\label{section:multi_edge}
In this section, we present an algorithm that takes a set $G$ of point-pairs that are edges in some triangulation of a lattice point-set and the Farey plans for the vectors of $G$ as inputs and outputs a flip plan that starts from an equilateral triangulation and forces $G$ to become edges.  
Moreover, we show that this is a minimum flip for $G$, and that all minimum flip plans for $G$ contain the same set of flips.  
This proves Theorem \ref{thm:main_flp} in Section \ref{sec:contributions}.  

\begin{figure}[htb]
    \centering
    \begin{subfigure}[t]{0.33\linewidth}
        \centering
        \includegraphics[width=0.5\linewidth]{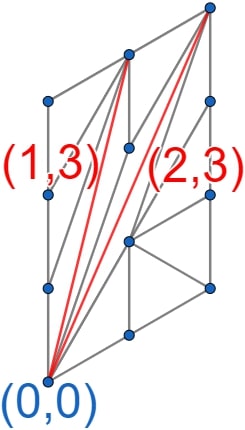}
        \caption{ }
        \label{fig:multiple_edges}
    \end{subfigure}
    \begin{subfigure}[t]{0.33\linewidth}
        \centering
        \includegraphics[width=1.1\linewidth]{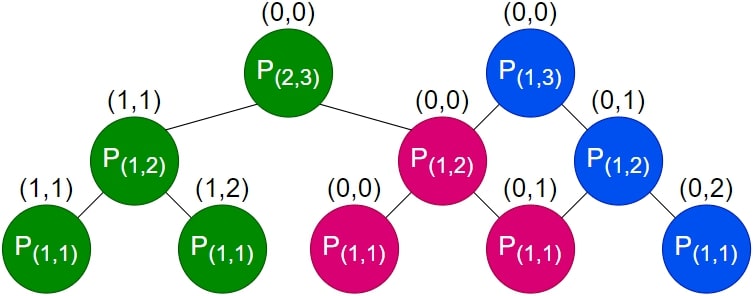}
        \caption{ }
        \label{fig:flip_plan_for_edges}
    \end{subfigure}
    \begin{subfigure}[t]{0.32\linewidth}
        \centering
        \includegraphics[width=0.7\linewidth]{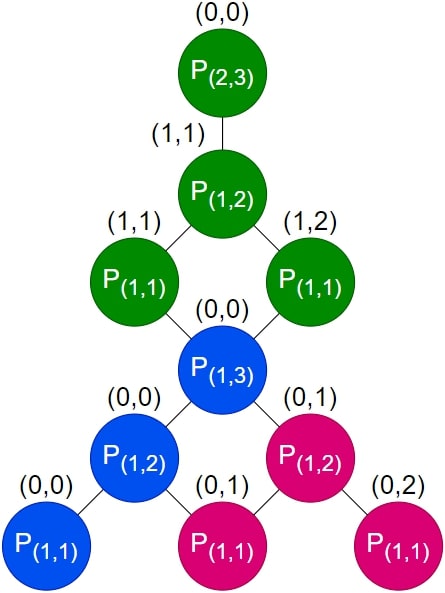}
        \caption{ }
        \label{fig:concatenation}
    \end{subfigure}
    \caption{(a) A triangulation containing the set $G = \{((1,3),(0,0)),((2,3),(0,0))\}$ of point-pairs as edges.  
    (b) A minimum flip plan $\pi_G$ for $G$ whose target triangulation is the one in (a).  
    The poset of blue and pink flips is the flip plan $\pi_{((1,3),(0,0))}$ and the poset of green and pink flips is the flip plan $\pi_{((1,3),(0,0))}$.  
    The poset of pink flips is $\pi_{((1,3),(0,0))} \cap \pi_{((2,3),(0,0))}$ (Definition \ref{def:intersection}) blue flips is $\pi_{((1,3),(0,0))} \setminus \pi_{((2,3),(0,0))}$, and green flips is $\pi_{((2,3),(0,0))} \setminus \pi_{((1,3),(0,0))}$.  
    (c) The poset $\pi_{((1,3),(0,0))} || \pi_{((2,3),(0,0))}$.  
    See Definition \ref{def:poset_diff_concat} and below.}
    \label{fig:multi_flip_plan}
\end{figure}

    \begin{note}
        From here on, unless otherwise specified, $\pi_{g,u}$ denotes the flip plan output by Algorithm Flip Plan on an input point-pair $(g,u)$.
    \end{note}
    
    \medskip\noindent
    \underbar{Algorithm \emph{Multi Flip Plan}} takes a set $G=\{(g_1,u_1),\dots,(g_n,u_n)\}$ of point-pairs and the Farey plans for the vectors of $G$ as inputs and outputs the poset of flips $\pi_{G} = \pi_{g_1,u_1} \cup \dots \cup \pi_{g_n,u_n}$.
    The algorithm uses a perfect hash table similar to the one in Algorithm Flip Plan to keep track of the poset computed thus far so that no redundant recursive calls or calls to Algorithm Flip Plan are made.
    
    \begin{theorem}[A Minimum Flip Plan for a Set of Point-pairs]
    \label{thm:MultiEdgeCreate}
      Let $G$ be a set of point-pairs that are edges in some triangulation of an equilateral lattice point-set.  
      On input $G$, Algorithm Multi Flip Plan outputs a minimum flip plan $\pi_G$ for $G$, and, if $G$ contains only distinct non-unit-length point-pairs, then this algorithm runs in $O(|\pi_G|)$ time.  
    \end{theorem}

    Theorem \ref{thm:MultiEdgeCreate} is proved at the end of this section using a corollary of Lemma \ref{lem:unique_parallelogram} along with Lemmas \ref{lem:farey_parallelogram_intersect}, \ref{lem:flip_plan_intersection}, and \ref{lem:flip_plan_for_independent}, below.  
    The following property of minimum flip plans is a byproduct of the proof of this theorem.  
    
    \begin{proposition}[All Minimum Flip Plans have the Same Set of Flips]
    \label{prop:MinFlipPlanSameSet}
    Let $G$ be a set of point-pairs that are edges in some triangulation of an equilateral lattice point-set. 
    All minimum flip plans for $G$ have the same set of flips.
    \end{proposition}

    \begin{figure}[htb]
        \centering
        \includegraphics[width=0.18\textwidth]{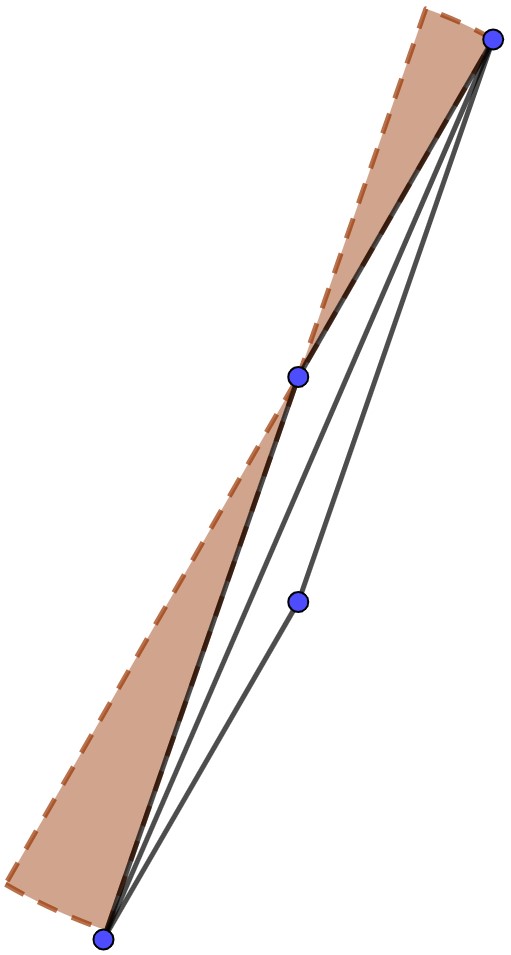}
        \caption{The Farey parallelogram and the regions that do not include a lattice point in the proof Lemma \ref{lem:unique_parallelogram}.}
        \label{fig:farey_parallelogram_regions}
    \end{figure}

\begin{lemma}[The Unique Flippable Parallelogram for a Point-Pair]
  \label{lem:unique_parallelogram}
    The Farey parallelogram for a point-pair is the unique parallelogram such that the region bounded by its point-pairs does not contain a lattice point.
  \end{lemma}

  \begin{proof}
    Consider a point-pair $(g,u)$.
    Corollary \ref{cor:EdgeCreate} shows that the Farey parallelogram $P_{g,u}$ does not contain a lattice point.  
    Consider any other parallelogram $P'_{g,u}$ and refer to the depiction of $P_{g,u}$ in Figure \ref{fig:farey_parallelogram_regions}.  
    Let the Farey parallelogram $P_g=\{g_1,g_2\}$, with $\phi_g(g_1) < \phi_g(g_2)$, and let $g'_1$ be the longer vector in the parallelogram $P'_g$.  
    If the region bounded by $P'_{g,u}$ does not contain a vertex of $P_{g,u}$, then a vertex of $P'_{g,u}$ must lie in one of the shaded circular arcs, excluding their boundaries.  
    This implies that either $\phi_g(g_1) < \phi_g(g'_1) < \phi_g(g)$ or $\phi_g(g) < \phi_g(g'_1) < \phi_g(g_2)$.  
    However, since the denominator of $\phi_g(g'_1)$ is less than the denominator of $\phi_g(g)$, one of the Farey neighbors of $\phi_g(g)$ in $F_g$ is not adjacent to $\phi_g(g)$ in $F_g$, which is a contradiction.  
    Thus, the region bounded by $P'_{g,u}$ contains a vertex of $P_{g,u}$, and the lemma is proved.
  \end{proof}
  
  \begin{corollary}[The Unique Flippable Quadrilateral for a Point-Pair]
  \label{cor:unique_quad}
  The Farey parallelogram for a point-pair $(g,u)$ is the unique quadrilateral for $(g,u)$ such that the region bounded by its point-pairs does not contain a lattice point.
  \end{corollary}
  
 This corollary shows that the Farey parallelogram for an edge is the minimal parallelogram defined in \cite{caputo2015}, and discussed in Section \ref{sec:intro_lat}.
 
 \begin{lemma}[Farey Parallelogram Intersection]
\label{lem:farey_parallelogram_intersect}
    Consider a non-unit-length point-pair $(g,u)$ and the flip plan $\pi_{g,u}$.  
    The shorter diagonal of the Farey parallelogram of any flip in $\pi_{g,u}$ and $(g,u)$ have intersecting line-segments.
\end{lemma}

The proof of Lemma \ref{lem:farey_parallelogram_intersect} is given in \ref{sec:proof_farey_parallelogram_intersect}
 
 \begin{lemma}[Point-pair Flip Plan Intersection]
\label{lem:flip_plan_intersection}
    Consider two point-pairs $(g,u)$ and $(g,v)$ and the flip plans $\pi_{g,u}$ and $\pi_{g,v}$.  
    If $\pi_{g,u}$ and $\pi_{g,v}$ both contain a flip on the Farey parallelogram for a point-pair $(h,y)$, then they both contain the flip plan $\pi_{h,y}$ as a subposet.
\end{lemma}

\begin{proof}
   The lemma follows by observing that, on inputs $(g,u)$ and $(g,v)$, Algorithm Flip Plan will recurse on $(h,y)$.  
\end{proof}
 
   We require the following definitions to state Lemma \ref{lem:flip_plan_for_independent}.
   
   \begin{definition}[Poset Difference and Concatenation]
   \label{def:poset_diff_concat}
Let $\pi$ and $\pi'$ be posets of flips.  The difference $\pi \setminus \pi'$ is the poset containing the flips in $\pi$ that are not contained in $\pi'$ such that two flips in $\pi \setminus \pi'$ are related if and only if they are related in $\pi$.  

The concatenation $\pi||\pi'$ is the poset $\pi \cup (\pi' \setminus \pi)$ with additional relations to ensure that each maximal flip in $\pi$ is performed before any flip in $\pi' \setminus \pi$.
\end{definition}

For example, see Figure \ref{fig:multi_flip_plan}. 

\begin{remark}[Complexity of Computing the Difference of Flip Plans]
\label{rem:diff_complexity}
In Section \ref{section:shortest_path}, we will want to quickly compute the difference $\pi' \setminus \pi$ of two posets of flips $\pi$ and $\pi'$.  
This can be achieved in $O(|\pi| + |\pi'|)$ time via a simple algorithm that (1) stores all flips in $\pi'$ in a perfect hash table $X$, similar to the one in Algorithm Flip Plan, and (2) outputting the poset obtained from $\pi$ by iterating through all its flips and removing those contained in $X$.
\end{remark}

\begin{definition}[Independent Set of Point-pairs]
\label{def:independent_point-pairs}
A set $G =  \{(g_1,u_1),\dots,$ \\$(g_n,u_n)\}$ of point-pairs is \emph{independent} if and only if each point-pair in $G$ is not unit-length and $\pi_{g_i,u_i}$ is not a subposet of $\pi_{g_j,u_j}$, for all $1 \leq i,j \leq n$ with $i \neq j$.
\end{definition}

Observe that any set $G$ of point-pairs has a unique maximal independent subset which, we call the \emph{maximum} independent subset of $G$.  
For example, if $G = \{((1,3)(0,0)),((2,3)(0,0)),((1,2)(0,0))\}$, then Figure \ref{fig:flip_plan_for_edges} shows that the maximum independent subset of $G$ is $\{((1,3)(0,0)),((2,3)(0,0))\}$.  

\begin{lemma}[A Flip Plan for Independent Point-pairs]
\label{lem:flip_plan_for_independent}
 Let $G = \{(g_1,u_1),\dots,$ \\$(g_n,u_n)\}$ be a set of point-pairs in an equilateral lattice point-set $S$ and let $G'$ be its maximum independent subset.  
  If $G'$ are edges in some triangulation of $S$, then $\tau_G = \pi_{g_1,u_1}||\dots||\pi_{g_n,u_n}$ is a flip plan for $G'$.
\end{lemma}

The proof of Lemma \ref{lem:flip_plan_for_independent} is given in Appendix \ref{sec:proof_flip_plan_for_independent}.  
We are now ready to prove Theorems \ref{thm:MultiEdgeCreate} and \ref{thm:main_flp}.  

\begin{proof}[Proof of Theorem \ref{thm:MultiEdgeCreate}]
The proof has 3 parts proving (1) the complexity of the algorithm, (2) that the output $\pi_G$ is a flip plan for $G$, and (3) that $\pi_G$  is a minimum flip plan.

    For (1), assume that $G$ contains only distinct non-unit-length point-pairs and let $(g,u)$ be the point-pair in $G$ that is currently under consideration in the algorithm.  
    Since the hash table is perfect, checking if it contains $(g,u)$ takes $O(1)$ time.  
    If the hash table contains $(g,u)$, then the algorithm moves onto the next point-pair in $G$, and so the total time spent on $(g,u)$ is $O(1)$.  
    Otherwise, $(g,u)$ is given as input to Algorithm Flip Plan.  
    
    Let $H$ be the subset of point-pairs in $G$ for which a call to Algorithm Flip Plan is made and let $H' = G \setminus H$.  
    Since $(g,u)$ is not unit-length, $\pi_{g,u}$ is non-empty.  
    Hence, $\pi_G$ contains a flip that adds $(g,u)$ as an edge.  
    Combining this with the fact that all edges in $G$ are distinct implies that $|H'| = O(|\pi_G|)$.  
    Therefore, the total time spent on point-pairs in $H'$ is $O(|\pi_G|)$.  
    Finally, a call to Algorithm Flip Plan is made for each point-pair in $H$, and each recursive call in this algorithm takes $O(1)$ time.  
    Observe that a recursive call adds a flip to $\pi_G$.  
    Thus, $O(|\pi_G|)$ total recursive calls are made, and the total run-time for Algorithm Multi Flip Plan on input $G$ is $O(|\pi_G|) + O(|\pi_G|)$, which proves the bound.  
    
    For (2), we first show that any consistent linear ordering $p = p_1\dots p_m$ of $\pi_G$ is a flip path.  
    Let $H=\{(h_1,y_1),\dots,(h_m,y_m)\}$ be the indexed set of point-pairs such that the flip $p_i$ adds the edge $(h_i,y_i)$.  
    Also, let $H'$ be the maximum independent subset of $H$.  
    Observe that $H'$ is a subset of $G$, and so it is a subset of edges of some triangulation, by assumption.  
    Hence, by Lemma \ref{lem:flip_plan_for_independent}, $\pi_H = \pi_{h_1,y_1}||\dots||\pi_{h_m,y_m}$ is a flip plan for $H'$.  
    By Lemma \ref{lem:flip_plan_intersection}, we see that $\pi_{h_i,y_i}$ is the subposet of $\pi_G$ whose maximal flip is $p_i$.  
    Consequently, since $p$ is a consistent linear ordering of $\pi_G$, we have $p_i = \pi_{h_i,y_i} \setminus (\pi_{h_1,y_1}||\dots||\pi_{h_{i-1},y_{i-1}})$.  
    Therefore, by definition, $\pi_H$ is the linear ordering $p$, and so $p$ is a flip path.
    
    Next, we show that $p$ is a flip path for $G$.  
    Consider any point-pair $(g_i,u_i)$ in $G \setminus H'$ and let $T'$ be the target triangulation of $p$.  
    If $(g_i,u_i)$ is unit-length, then it is contained in the starting triangulation of $p$.  
    Otherwise, by the definition of $H'$, some flip in $p$ adds $(g_i,u_i)$ as an edge.  
    Furthermore, since the line segment of $(g_i,u_i)$ does not intersect the line segment of any other point-pair in $G$, no flip in $p$ removes $(g_i,u_i)$, by Lemma \ref{lem:farey_parallelogram_intersect}.  
    Therefore, the target triangulation of $p$ contains $G$ as edges.
    
    For (3), since $\pi_G$ is a flip plan, it suffices to show that some shortest flip path contains all the flips in $\pi_G$.  
    We prove a stronger statement that \emph{every} shortest flip path  contains all the flips in $\pi_G$, which proves Proposition \ref{prop:MinFlipPlanSameSet}.  
    Let $q=q_1,\dots,q_r$ be any shortest flip path for $G$.  
    If $\pi_G$ is empty, then $q$ is empty and the theorem is proved.  
    Otherwise, $G$ contains at least one non-unit-length edge.  
    
    \emph{Claim:} all flips in $q$ replace edges with longer ones.  
    
    Assuming the claim, $q$ contains all maximal flips in $\pi_G$, by Corollary \ref{cor:unique_quad}.  
    Let $\pi'_G$ be $\pi_G$ with its maximal flips removed and consider any maximal flip in $\pi_G$, and let it be performed on the Farey parallelogram $P_{g_i,u_i}$.  
    If the longer point-pairs in $P_{g_i,u_i}$ are not unit-length, then $q$ must contain flips on their Farey parallelograms.  
    These are the maximal flips in $\pi'_G$, by Lemma \ref{lem:flp_neighbor}.  
    By repeating this argument, we see that $q$ contains all flips in $\pi_G$.  
    Since $\pi_G$ is a flip plan and $q$ is a shortest flip path, $q$ does not contain any other flips.  
    This proves Theorem \ref{thm:MultiEdgeCreate} and Proposition \ref{prop:MinFlipPlanSameSet}. 
    
    We conclude the proof by proving the above claim.  
    Assume that some flip in $q$ replaces an edge with a shorter one and let $q_t$ be the first such flip, which replaces the edge $(h,y)$.  
    Since $(h,y)$ is not unit-length, the Farey parallelogram $P_{h,y}$ exists, and $q_t$ is a flip on $P_{h,y}$, by Corollary \ref{cor:unique_quad}.  
    Also, since all edges in the starting triangulation of $q$ are unit-length, we have $t > 1$.  
    Hence, by the same corollary, some flip $q_s$, with $1 \leq s < t$, is on $P_{h,y}$ and adds $(h,y)$ as an edge.  
    The discussion above implies that the triangulations resulting from $q_s$ and $q_{t-1}$ contain the point-pairs in $P_{h,y}$ as edges.  
    Therefore, since each flip between $q_s$ and $q_t$ replace an edge with a longer one, by assumption, the triangulation resulting from this flip also contains the point-pairs in $P_{h,y}$ as edges.  
    If neither $(h,y)$ nor the shorter diagonal of $P_{h,y}$ is a point-pair in $G$, then removing $q_s$ and $q_t$ from $q$ yields a shorter flip path between the equilateral triangulation and $T''$.  
    Otherwise, we can obtain a similar outcome by removing either $q_s$ from $q$ if the shorter diagonal of $P_{h,y}$ is a point-pair in $G$, or $q_t$ from $q$ if $(h,y)$ is a point-pair in $G$.  
    Thus, in all cases we contradict the fact that $q$ is a shortest flip path, proving the claim.  
\end{proof}

\begin{proof}[Proof of Theorem \ref{thm:main_flp}]
   If the given point-pair is unit-length, then the fact that its flip plan is minimum follows from Theorem \ref{thm:MultiEdgeCreate} and the complexity statement follows from Corollary \ref{cor:EdgeCreate}.  
   Otherwise, the theorem follows from Theorem \ref{thm:MultiEdgeCreate}.
\end{proof}

\section{Minimum Flip Plans Starting from Any Minimum Triangulation}
\label{sec:ground_state}
In this section, we generalize Theorem \ref{thm:MultiEdgeCreate} by allowing the starting triangulation to be any \emph{minimum triangulation} (Definition \ref{def:minimum_tri}).  
Then, we show that any triangulation is a minimum triangulation containing a set of edges, as discussed in Section \ref{sec:contributions}.  

\begin{note}
    From here on, unless otherwise specified, $\pi_G$ denotes the flip plan output by Algorithm Multi Flip Plan on an input set $G$ of point-pairs.  
    
    Additionally, consider a lattice point-set $S = L \cap \Omega$.  
    We can treat the boundary line-segments of the polygon $\Omega$ as a set of point-pairs.  
    We denote this set by $\Omega$, and it will be clear from context whether we are referring to a polygon or a set of point-pairs.  
\end{note}

We begin with definitions and then state the main theorem of this section.

\begin{figure}[htb]
    \centering
    \begin{subfigure}[t]{0.24\linewidth}
        \centering
        \includegraphics[width=0.8\linewidth]{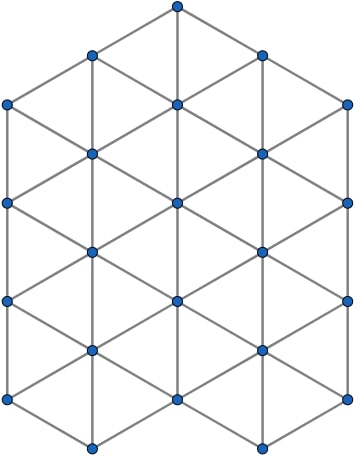}
        \caption{}
    \end{subfigure}
    \begin{subfigure}[t]{0.24\linewidth}
        \centering
        \includegraphics[width=0.8\linewidth]{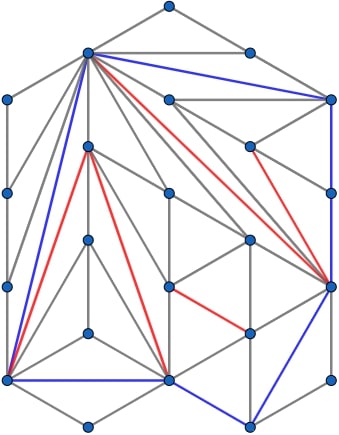}
        \caption{}
    \end{subfigure}
    \begin{subfigure}[t]{0.24\linewidth}
        \centering
        \includegraphics[width=0.8\linewidth]{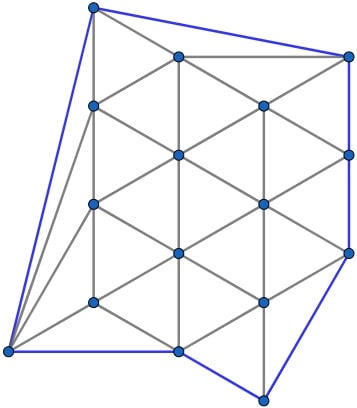}
        \caption{}
    \end{subfigure}
    \begin{subfigure}[t]{0.24\linewidth}
        \centering
        \includegraphics[width=0.8\linewidth]{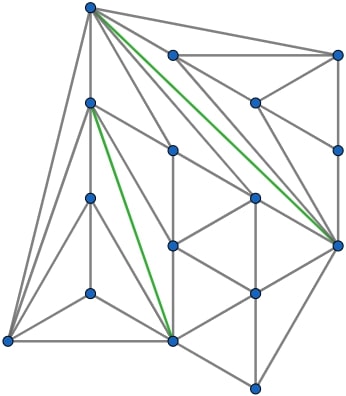}
        \caption{}
    \end{subfigure}
    \caption{(a) and (c) Minimum triangulations (Definition \ref{def:minimum_tri}).  
    (b) A minimum triangulation containing the set of blue and red point-pairs as edges.  
    (d) A minimum triangulation containing the set of green point-pairs as edges.  
    The blue edges from (b) are the polygonal boundary edges in (c) and (d).  
    See below.  }
    \label{fig:unique_tri}
\end{figure}

\begin{definition}[Point-pairs that are independent in a Particular Lattice Point-set]
\label{def:ind_in_S}
A set $G=\{(g_1,u_1),\dots,(g_n,u_n)\}$ of point-pairs in a lattice point-set $S=L \cap \Omega$ is \emph{independent in $S$} if and only if $G$ does not contain any point-pair in $\Omega$ and the maximum independent subset of $G \cup \Omega$ contains $G$.
\end{definition}

For any set of point-pairs $G$ and any lattice point-set $S$, note that $G$ has a unique maximal subset that is independent in $S$, which we call the \emph{maximum subset of $G$ that is independent in $S$}.  
If $S$ admits an equilateral triangulation, then $G$ is independent in $S$ if and only if it is independent as in Definition \ref{def:independent_point-pairs}.  
Additionally, the \emph{maximum independent set} of a triangulation $T$ of $S$ is the maximum subset of edges in $T$ that is independent in $S$.  

For example, consider Figure \ref{fig:unique_tri} and let $G$ be the set of red edges in (b), $G'$ be the set of green edges in (d), $\Omega$ be the set of blue edges in (b)-(d), and $\Omega'$ be the subset of all non-unit-length edges in $\Omega$.  
Also, let $S$ be the lattice point-set in (a) and (b) and let $S'$ be the lattice point-set in (c) and (d).  
The set $G' \cup \Omega'$ is both the maximum independent subset of $G \cup \Omega$ and the maximum subset of $G \cup \Omega$ that is independent in $S$.  
The set $G'$ is the maximum subset of $G \cup \Omega$ that is independent in $S'$.  

\begin{definition}[Minimum Triangulations]
\label{def:minimum_tri}
    Let $G$ be a set of point-pairs that are edges in some triangulation of a lattice point-set $S$ and let $G'$ be its maximum subset that is independent in $S$.
    \begin{itemize}
        \item A \emph{minimum triangulation}, denoted by $MT(S)$, is a triangulation such that each of its edges is either unit-length or some point-pair in the Farey parallelogram for the edge is not an edge in $MT(S)$.
    
        \item A minimum triangulation \emph{containing $G$ as edges}, denoted by $MT\left(S,G\right)$, is a triangulation containing $G$ as edges such that $G'$ is the set of all edges for which the point-pairs in their Farey parallelograms are edges in $MT\left(S,G\right)$.  
        If $G'$ is empty, then $MT(S,G)=MT(S)$.
    \end{itemize}
\end{definition}

See Figure \ref{fig:unique_tri} for examples of these triangulations.  
Lemma \ref{lem:unique_minimum_tri}, below, shows that these triangulations are unique.
Given their uniqueness, we can state the main theorem of this section as follows.  
Recall that a set of constraint edges is a set of point-pairs that must be edges in all triangulations along a flip path.  

\begin{theorem}[Constrained Minimum Flip Plans that Start from Minimum Triangulations]
\label{thm:constrained_flp_not_equilateral}
    Let $F$ be a set of constraint edges and let $G$ be a set of point-pairs in a lattice point-set $S=L \cap \Omega$ such that $F \cup G$ are edges in some triangulation of $S$.
    Also, consider the flip plans $\pi_{\Omega}$, $\pi_F$, and $\pi_G$.  
    Then, the following statements are true.
    \begin{enumerate}
        \item $\pi_G \setminus (\pi_{\Omega} \cup \pi_F)$ is a minimum flip plan constrained by $F$ that starts from the minimum triangulation $MT(S,F)$ and forces $G$ to become edges,
        \item $\pi_G \setminus (\pi_{\Omega} \cup \pi_F)$ is a minimum flip plan constrained by $F$ between $MT(S,F)$ and the minimum triangulation $MT(S,F \cup G)$, and
        \item  all minimum flip plans of the above types have the same set of flips.
    \end{enumerate}
\end{theorem}

The theorem is proved using Lemmas \ref{lem:unique_minimum_tri} and \ref{lem:constrained_flp_equilateral}, below.  
The proofs of these lemmas are given in \ref{sec:proof_unique_minimum_tri} and \ref{sec:proof_constrained_flp_equilateral}, respectively.  

\begin{lemma}[Unique Minimum Triangulations]
\label{lem:unique_minimum_tri}
    Consider a set $G$ of point-pairs that are edges in some triangulation of a lattice point-set $S$.  
    The minimum triangulation of $S$ containing $G$ as edges is unique.
\end{lemma}

The proof of Lemma \ref{lem:unique_minimum_tri} goes as follows.  
First, we prove the Lemma when $S = L \cap \Omega$ admits an equilateral triangulation (Lemma \ref{lem:flp_target_minimum}).  
In the case where $S$ does not admit such a triangulation, we consider a point-set $S'$ that contains $S$ and which does admit an equilateral triangulation.  
Then, we show that removing the points in $S' \setminus S$ from the unique minimum triangulation $MT(S', \Omega \cup G)$ yields a minimum triangulation $MT(S,G)$.  
Finally, we prove that $MT(S,G)$ is unique by showing that distinct triangulations of $S$ that contain $G$ as edges correspond to distinct minimum triangulations of $S'$ that differ in their edges between points in $S$.  
As an illustration of this proof, refer to the minimum triangulations in Figure \ref{fig:unique_tri} (b) and (d) and let $G$ be the red edges and $\Omega$ be the blue edges.  
The set of green edges is the maximum subset of $G$ that is independent in point-set in Figure \ref{fig:unique_tri}(d).  

\begin{lemma}[Constrained Minimum Flip Plans Starting from an Equilateral triangulation]
\label{lem:constrained_flp_equilateral}
    Let $F$ be a set of constraint edges and a let $G$ be a set of point-pairs in a lattice point-set $S$ such that $F \cup G$ is a subset of edges of some triangulation of $S$.
    Also, consider the flip plans $\pi_F$ and $\pi_G$.  
    Then, the following statements are true.
    \begin{enumerate}
        \item $\pi_G \setminus \pi_F$ is a minimum flip plan constrained by $F$ that starts from the minimum triangulation $MT(S,F)$ and forces $G$ to become edges,
        \item $\pi_G \setminus \pi_F$ is a minimum flip plan constrained by $F$ between $MT(S,F)$ and the minimum triangulation $MT(S,F \cup G)$, and
        \item all minimum flip plans of the above types have the same set of flips.
    \end{enumerate}
\end{lemma}

To prove Lemma \ref{lem:constrained_flp_equilateral}, we use the above-mentioned Lemma \ref{lem:flp_target_minimum} to show that if any of the lemma statements is false, then so is Theorem \ref{thm:MultiEdgeCreate}, which is a contradiction.  

We are now ready to prove the Theorem \ref{thm:constrained_flp_not_equilateral}.

\begin{proof}[Proof of Theorem \ref{thm:constrained_flp_not_equilateral}]
   Consider a lattice point-set $S'$ that contains $S$ and admits an equilateral triangulation.  
   Since $F \cup G$ is a subset of edges of some triangulation of $S$, and all triangulations of $S$ contain $\Omega$ as edges, $\Omega \cup F \cup G$ is a subset of edges of some triangulation of $S'$.  
   Hence, $\pi_G \setminus (\pi_{\Omega} \cup \pi_F)$ is a minimum flip plan constrained by $\Omega \cup F$ that starts from the minimum triangulation $MT(S',\Omega \cup F)$ and forces $G$ to become edges, by Lemma \ref{lem:constrained_flp_equilateral}.  
   Additionally, the target triangulation of $\pi_G \setminus (\pi_{\Omega} \cup \pi_F)$ is the minimum triangulation $MT(S',\Omega \cup F \cup G)$.  
   
   Next, as in the proof of Lemma \ref{lem:unique_minimum_tri} (see \ref{sec:proof_unique_minimum_tri} and above proof sketch), removing the points in $S' \setminus S$ from these triangulations yields the minimum triangulations $MT(S,F)$ and $MT(S,F \cup G)$.  
   We will show that all flips in $\pi_G \setminus (\pi_{\Omega} \cup \pi_F)$ are on Farey parallelograms contained in $S$.  
   This demonstrates that $\pi_G \setminus (\pi_{\Omega} \cup \pi_F)$ is a flip plan constrained by $F$ that starts from the $MT(S,F)$ and forces $G$ to become edges.  
   Moreover, its target triangulation is $MT(S,F \cup G)$.  
   
   Assume to the contrary that some flip in $\pi_G \setminus (\pi_{\Omega} \cup \pi_F)$ is on a Farey parallelogram $P_{h,y}$, for a point-pair $(h,y)$, with a vertex in $S' \setminus S$.  
   Since each point-pair in $G$ is between points in $S$, this implies that the line-segments of a point-pair in $P_{h,y}$ and a point-pair in $\Omega$ intersect.  
   Hence, these point-pairs cannot both be edges in the same triangulation.  
   However, as noted above, $\pi_G \setminus (\pi_{\Omega} \cup \pi_F)$ is constrained by $\Omega \cup F$, and so this is a contradiction.
   
   Finally, if $\pi_G \setminus (\pi_{\Omega} \cup \pi_F)$ is not a minimum flip plan as described by either Statement (1) or (2), then we can easily obtain a contradiction as in the proof of Lemma \ref{lem:constrained_flp_equilateral}.  
\end{proof}

We conclude this section with the following proposition.  

\begin{proposition}[Identity Map Between Triangulations and Minimum Triangulations]
\label{prop:bij_tri_min_2}
    Let $T$ be a triangulation of a lattice point-set $S$ and let $G'$ be its maximum independent set.  
    Then, the map that takes $T$ to the the minimum triangulation of $S$ containing $G'$ as edges is the identity map.  
\end{proposition}

\begin{proof}
    Let $S=L \cap \Omega$ and let $G$ be the set of all edges in $T$.
    It suffices to show that the composition of the map that sends $T$ to $G'$ and the map that sends $G'$ to the minimum triangulation $MT(S,G')$ is the identity map.  
    First, observe that these maps are well-defined.  
    Second, by Statements (1) of Theorem \ref{thm:constrained_flp_not_equilateral}, $T$ is the target triangulation of $\pi_G \setminus \pi_{\Omega}$.  
    Furthermore, by Statement (2) of this theorem, $T$ is the minimum triangulation $MT(S,G)$.  
    Finally, by definition, $MT(S,G)$ is $MT(S,G')$.  
    Thus, the second map sends $G'$ to $T$.  
\end{proof}

\section{A Constrained Minimum Flip Plan Between Two Triangulations}
\label{section:shortest_path}
In this section, we prove Theorems \ref{thm:main_fp} and \ref{thm:main_rectangular_complexity}, stated in Section \ref{sec:contributions}.  
Theorem \ref{thm:main_fp} follows from Theorem \ref{thm:constr_min_flp} and Lemma \ref{lem:applying_thm_to_tris}, below.  
We require the following definitions to state and prove Theorem \ref{thm:constr_min_flp}.  

\begin{figure}[htb]
    \centering
    \begin{subfigure}[t]{0.32\linewidth}
        \centering
        \includegraphics[width=0.5\linewidth]{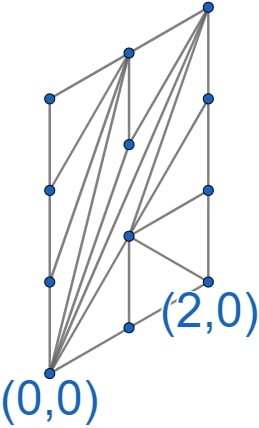}
        \caption{}
    \end{subfigure}
    \begin{subfigure}[t]{0.32\linewidth}
        \centering
        \includegraphics[width=0.5\linewidth]{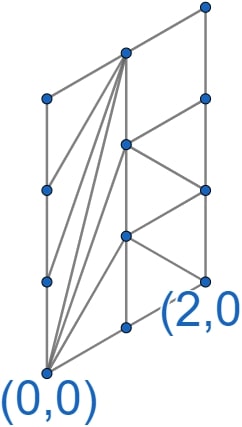}
        \caption{}
    \end{subfigure}
    \begin{subfigure}[t]{0.32\linewidth}
        \centering
        \includegraphics[width=0.5\linewidth]{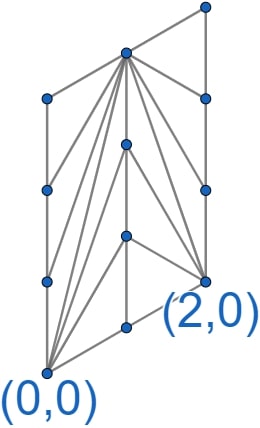}
        \caption{}
    \end{subfigure}
    \begin{subfigure}[t]{0.32\linewidth}
        \centering
        \includegraphics[width=1.2\linewidth]{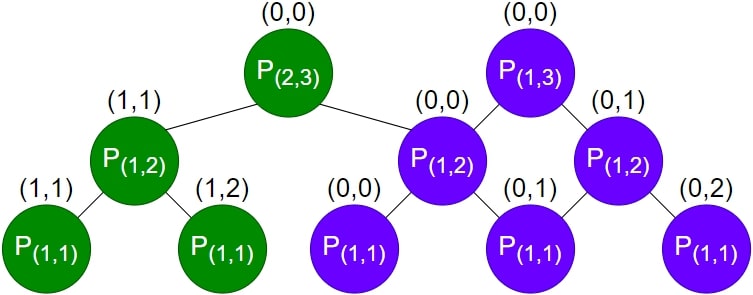}
        \caption{}
    \end{subfigure}
    \begin{subfigure}[t]{0.32\linewidth}
        \centering
        \includegraphics[width=0.45\linewidth]{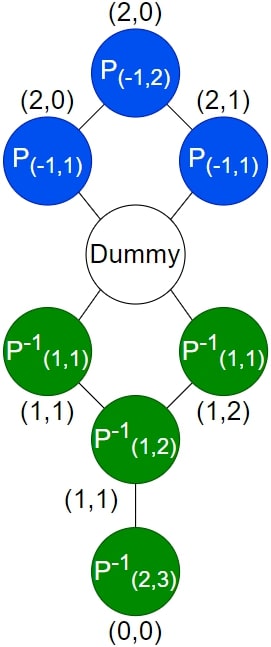}
        \caption{}
    \end{subfigure}
    \begin{subfigure}[t]{0.24\linewidth}
        \centering
        \includegraphics[width=1.5\linewidth]{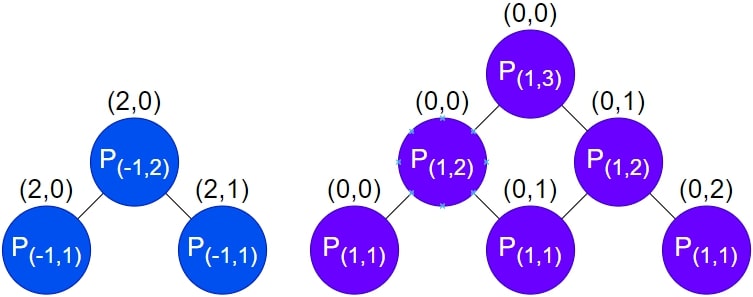}
        \caption{}
    \end{subfigure}
    \caption{(d) and (f) Minimum flip plans $\pi_1$ and $\pi_2$ that start from an equilateral triangulation and end with the triangulations in (a) and (c), respectively.  
    The poset $\pi_1 \cap \pi_2$ (Definition \ref{def:intersection}) is shown in purple, $\pi_1 \setminus \pi_2$ is shown green, and $\pi_2 \setminus \pi_1$ is shown in blue.  
    (e) The minimum flip plan $\pi^{-1}_1 || \pi_2$ between the triangulations in (a) and (c), where concatenation is handled using a dummy flip (see proof of Proposition \ref{prop:overall_complexity}).  
    Inverse flips are denoted by a superscript ``$-1$.''  
    (b) The target triangulation of $\pi^{-1}_1$.  
    See Definition \ref{def:inverse} and below.}
    \label{fig:constr_flip_plan}
\end{figure}

\begin{definition}[Poset Intersection]
    \label{def:intersection}
    The intersection $\pi \cap \pi'$ of two posets $\pi$ and $\pi'$ of flips is the maximum-size poset contained in both $\pi$ and $\pi'$.  
\end{definition}

For example, see Figure \ref{fig:multi_flip_plan}.  

\begin{definition}[Inverse Flip and Inverse Flip Plan]
\label{def:inverse}
The \emph{inverse} of a flip on a Farey parallelogram $P_{g,u}$ that adds the point-pair $(g,u)$ as an edge is the flip on $P_{g,u}$ that replaces the edge $(g,u)$ with the shorter diagonal of $P_{g,u}$.  

The \emph{inverse} $\pi^{-1}$ of a flip plan $\pi$, whose flips are on Farey parallelograms and replace edges with longer ones, is obtained from $\pi$ be reversing all relations in its partial order and inverting all of its flips.
\end{definition}

For example, consider the flip plans $\pi_{(1,6),(0,0)}$ and $\pi_{(3,5),(0,0)}$ in Figure \ref{fig:more_flip_plans}.  
The inverses of these flip plans contain inverse flips on the Farey parallelograms $P_{(1,6),(0,0)}$ and $P_{(3,5),(0,0)}$ as minimal elements.  
Furthermore, the union of these inverse flip plans is a flip plan whose starting triangulation is the one in Figure \ref{fig:fp_7} and whose target triangulation is the equilateral triangulation in Figure \ref{fig:fp_1}.

\begin{theorem}[Constrained Minimum Flip Plan Between Minimum Triangulations]
\label{thm:constr_min_flp}
    Let $F$ be a set of constraint edges and let $G \cup F$ and $G' \cup F$ be sets of point-pairs that are each edges in some triangulation of a lattice point-set $S=L \cap \Omega$.  
    Then, the poset
    $$(\pi_G \setminus (\pi_{\Omega} \cup \pi_F \cup \pi_{G'}))^{-1} || (\pi_{G'} \setminus (\pi_{\Omega} \cup \pi_F \cup \pi_G)),$$ 
    is a minimum flip plan constrained by $F$ between the minimum triangulations $MT(S,F \cup G)$ and $MT(S,F \cup G')$.  
    Furthermore, all minimum flip plans between these triangulations contain the same set of flips.  
\end{theorem}

\begin{proof}
    Let $\pi$ be the poset given by the theorem and consider the flip plans $\pi_{\Omega}$, $\pi_F$, $\pi_G$, $\pi_{G'}$, and $\pi_H = \pi_G \cap \pi_{G'}$, where $H$ is the set of point-pairs added by the maximal flips in this intersection.  
    First, we show that $\pi$ is a flip plan between the minimum triangulations $MT(S,F \cup G)$ and $MT(S,F \cup G')$.  
    By the Theorem \ref{thm:constrained_flp_not_equilateral}, $\pi_G \setminus (\pi_{\Omega} \cup \pi_F)$ is a minimum flip plan constrained by $F$ between the minimum triangulations $MT(S,F)$ and $MT(S,F \cup G)$.  
    Similarly, $\pi_{G'} \setminus (\pi_{\Omega} \cup \pi_F)$ is a minimum flip plan constrained by $F$ between $MT(S,F)$ and $MT(S,F \cup G')$.  
    Since $\pi_H$ is a subposet of both $\pi_G$ and $\pi_{G'}$, $\pi_H \setminus (\pi_{\Omega} \cup \pi_F)$ is a minimum flip plan constrained by $F$ between $MT(S,F)$ and the minimum triangulation $MT(S,F \cup H)$.  
    Hence, we get that $\pi_{G'} \setminus (\pi_{\Omega} \cup \pi_F \cup \pi_H)$ is a minimum flip plan constrained by $F$ between $MT(S,F \cup H)$ and $MT(S,F \cup G')$.  
    Similarly, $\pi_{G'} \setminus (\pi_{\Omega} \cup \pi_F \cup \pi_H)$ is a minimum flip plan constrained by $F$ between $MT(S,F \cup H)$ and $MT(S,F \cup G')$.  
    The fact that $\pi$ is a flip plan between $MT(S,F \cup G)$ and $MT(S,F \cup G')$ follows easily from the observations that $\pi_G \setminus \pi_{G'} = \pi_G \setminus \pi_H$ and $\pi_{G'} \setminus \pi_G = \pi_{G'} \setminus \pi_H$.  
    
    Next, we prove that $\pi$ is a minimum flip plan that starts from $MT(S,F \cup G)$ and ends with $MT(S,F \cup G')$ by demonstrating that any shortest flip path $p=p_1,\dots,p_n$ that starts from $MT(S,F \cup G)$ and ends with $MT(S,F \cup G')$ contains all flips in $\pi$.  
    It suffices to show that $p$ contains all flips in $\pi' = (\pi_G \setminus (\pi_{\Omega} \cup \pi_F \cup \pi_{G'}))^{-1}$.  
    If this is true, then swapping the roles of $G$ and $G'$ shows that $p$ contains all flips in $\pi_{G'} \setminus (\pi_{\Omega} \cup \pi_F \cup \pi_G$, and hence $\pi$.  
    This proves the theorem.  
    
    We now complete the proof by showing that $p$ contains all flips in $\pi'$.  
    If $\pi'$ is empty, then we are done.  
    Otherwise, let $s^{-1}$ be any inverse flip in $\pi'$, where $s^{-1}$ is the inverse of some flip $s$ in $\pi_G \setminus (\pi_{\Omega} \cup \pi_F \cup \pi_{G'})$.  
    Also, consider the triangulations $T_0,\dots,T_n$ resulting from each flip in $p$ and let $\pi_i$ be the minimum flip plan that starts from the minimum triangulation $MT(S)$ and forces the maximum independent set of $T_i$ to become edges, given by Theorem \ref{thm:constrained_flp_not_equilateral}.  
    As in the proof of Proposition \ref{prop:bij_tri_min_2}, $T_i$ is the target triangulation of $\pi_i$.  
    Hence, it is easy to see that each flip in $\pi_G \setminus (\pi_{\Omega} \cup \pi_F \cup \pi_{G'})$, including $s$, is contained in $\pi_0$ but not $\pi_n$.  
    Additionally, note that $\pi_i$ and $\pi_{i+1}$ differ by a single maximal flip, for all $0 \leq i \leq n-1$.  
    These facts imply the existence of an integer $0 \leq j \leq n-1$ such that $s$ is contained in $\pi_j$ but not $\pi_{j+1}$ and $\pi_{j+1}$ can be obtained from $\pi_j$ be deleting $s$.  
    This implies that $s^{-1}$ is the unique flip that transforms $T_j$ into $T_{j+1}$.  
    Thus, $p$ contains $s^{-1}$, and the proof is complete.  
\end{proof}

\begin{lemma}[Constrained Minimum Flip Plan Between Triangulations]
\label{lem:applying_thm_to_tris}
    Let $T$ and $T'$ be any two triangulations of a lattice point-set whose edge sets are $G$ and $G'$, respectively, and whose maximum independent sets are $H$ and $H'$, respectively.  
    Also, let $F$ be a set of constraint edges of both $T$ and $T'$.  
    Then, the minimum flip plan obtained by applying Theorem \ref{thm:constr_min_flp} to either $F$, $G$, and $G'$ or $F$, $H$, and $H'$ is a minimum flip plan constrained by $F$ between $T$ and $T'$.  
\end{lemma}

\begin{proof}
    By Proposition \ref{prop:bij_tri_min_2}, $T$ and $T'$ are the minimum triangulations $MT(S,H)$ and $MT(S,H')$, respectively.  
    Also, by definition, we have $MT(S,H) = MT(S,F \cup H) = MT(S,F \cup G)$ and $MT(S,H') = MT(S,F \cup H') = MT(S,F \cup G')$.  
    Thus, the lemma follows by applying Theorem \ref{thm:constr_min_flp} to either $F$, $G$, and $G'$ or $F$, $H$, and $H'$.  
\end{proof}

\begin{proof}[Proof of Theorem \ref{thm:main_fp}]
    The theorem follows immediately from applying Lemma \ref{lem:applying_thm_to_tris} to the given triangulations.   
\end{proof}

Next, Proposition \ref{prop:output_sensitive_complexity} gives a large class of pairs of triangulations such that the constrained minimum flip plan between them can be computed in time linear its total number of flips.  
The proof of this proposition requires Proposition \ref{prop:overall_complexity}, below.  

\begin{proposition}[Overall Complexity of Computing Constrained Minimum Flip Plans]
\label{prop:overall_complexity}
Consider a lattice point-set $S = L \cap \Omega$ and a set $F$ of constraint edges.  
Given the maximum independent subset $\Omega'$ of point-pairs in $\Omega$ and the maximum independent sets $G$ and $G'$ of two triangulations $T$ and $T'$ of $S$, respectively, a constrained minimum flip plan between $T$ and $T'$ can be computed in $O(|\pi_{\Omega}| + |\pi_G| + |\pi_{G'}|)$ time.
\end{proposition}

\begin{proof}
    By Lemma \ref{lem:applying_thm_to_tris}, the minimum flip plan $\pi$ obtained by applying Theorem \ref{thm:constr_min_flp} to $F$, $G$, and $G'$ is a constrained minimum flip plan between $T$ and $T'$.  
    Hence, it suffices to show that we can compute $\pi$ in the desired time.  
    By definition, $\Omega'$, $G$, and $G'$ are each a set of distinct non-unit-lengths edges.  
    Consequently, by Theorem \ref{thm:MultiEdgeCreate}, the flip plans $\pi_{\Omega'}$, $\pi_G$, and $\pi_{G'}$ can be computed in $O(|\pi_{\Omega'}| + |\pi_G| + |\pi_{G'}|)$ time.  
    Observe that $\pi_{\Omega} = \pi_{\Omega'}$ and $\pi_F$ is a subposet of both $\pi_G$ and $\pi_{G'}$.  
    Therefore, $|\pi_{\Omega}| = |\pi_{\Omega'}|$ and we can rewrite $\pi$ as
    \begin{equation*}
        (\pi_G \setminus (\pi_{\Omega} \cup \pi_{G'}))^{-1} || (\pi_{G'} \setminus (\pi_{\Omega} \cup \pi_G)).
    \end{equation*}
    The poset unions and inversion in $\pi$ can clearly be computed in $O(|\pi_{\Omega}| + |\pi_G| + |\pi_{G'}|)$ time.  
    Additionally, as discussed in Remark \ref{rem:diff_complexity} in Section \ref{section:multi_edge}, the poset differences in $\pi$ can also be computed in this time.  
    Lastly, poset concatenation can be handled in the desired time by adding a dummy flip to $\pi$ that is the parent of all maximal flips in $(\pi_G \setminus (\pi_{\Omega} \cup \pi_{G'}))^{-1}$ and the child of all minimal flips in $\pi_{G'} \setminus (\pi_{\Omega} \cup \pi_G)$.  
    This completes the proof.  
\end{proof}

\begin{proposition}[Output Sensitive Complexity]
\label{prop:output_sensitive_complexity}
Let $T$ and $T'$ be triangulations of a lattice point-set $S = L \cap \Omega$ whose maximum independent sets are $G$ and $G'$, respectively.  
If $\pi_{\Omega}$ and $\pi_G \cap \pi_{G'}$ are empty, then a constrained minimum flip plan $\pi$ between $T$ and $T'$ can be found in $O(|\pi|)$ time.  
\end{proposition}

\begin{proof}
    Assume that $\pi_{\Omega}$ and $\pi_G \cap \pi_{G'}$ are empty and let $F$ be a set of constraint edges of both $T$ and $T'$.  
    By Lemma \ref{lem:applying_thm_to_tris}, the minimum flip plan $\pi$ obtained by applying Theorem \ref{thm:constr_min_flp} to $F$, $G$, and $G'$ is a minimum flip plan constrained by $F$ between $T$ and $T'$.  
    Hence, it suffices to show that $\pi$ can be computed in $O(|\pi|)$ time.  
    
    By our assumptions, $\pi$ can be rewritten as $\pi_G^{-1} || \pi_{G'}$.  
    Since $G$ and $G'$ each contain only distinct non-unit-length edges, by definition, Theorem \ref{thm:MultiEdgeCreate} tells us that $\pi_G$ and $\pi_{G'}$ can be computed in $O(|\pi_G| + |\pi_{G'}|)$ time, and hence in $O(|\pi|)$ time.  
    Finally, the poset union, inversion, and concatenation in $\pi$ can be computed in $O(|\pi|)$ time, as noted in the proof of Proposition \ref{prop:overall_complexity}.  
    This proves the proposition.  
\end{proof}

Finally, we prove Theorem \ref{thm:main_rectangular_complexity}.  

\begin{proof}[Proof of Theorem \ref{thm:main_rectangular_complexity}]
    Let $T$ and $T'$ be triangulations of an $n$-point rectangular lattice point-set $S = L \cap \Omega$.  
    Since a triangulation of any point-set with $n$ points contains $O(n)$ edges, for any standard representation of a triangulation (e.g., a doubly connected edge list), it is easy to see that we can obtain the sets $G$ and $G'$ of all edges in $T$ and $T'$, respectively, represented as point-pairs in a three-direction lattice in $O(n)$ time.  
    Let $F$ be a set of constraint edges contained in both $G$ and $G'$ and let $H$ and $H'$ be the maximal independent sets of $T$ and $T'$, respectively.  
    By Lemma \ref{lem:applying_thm_to_tris}, the minimum flip plan $\pi$ obtained by applying Theorem \ref{thm:constr_min_flp} to $F$, $H$, and $H'$ is a minimum flip plan constrained by $F$ between $T$ and $T'$.  
    Therefore, it suffices to show that we can compute $\pi$ in $O(n^{\frac{3}{2}})$ time.  
    
    By definition, the edges in $H$ and $H'$ are the only ones in either $T$ or $T'$ such that the point-pairs in their Farey parallelograms are also edges.  
    Hence, since $G$ and $G'$ contain all edges in $T$ and $T'$, respectively, it is easy to see that we can compute $H$ and $H'$ in $O(n)$ time.  
    Furthermore, since $S$ is rectangular, $\Omega$ contains only unit-length edges.  
    This implies that the maximum independent subset of point-pairs in $\Omega$ is empty.  
    Therefore, by Proposition \ref{prop:overall_complexity}, $\pi$ can be computed in $O(|\pi_H| + |\pi_{H'}|)$ time.  
    Finally, since $\pi_H$ and $\pi_{H'}$ are both flip plans that start from the equilateral triangulation of $S$, Lemma \ref{lem:caputo_bound} tells us that $|\pi_H| + |\pi_{H'}| = O(n^\frac{3}{2})$, which proves the theorem.
\end{proof}

\section{Conclusion and Further Works}
\label{section:conclusion}
    We have elucidated the structure of constrained, shortest flip paths between lattice triangulations, given algorithms with improved complexity over previous algorithms, and given output-sensitive algorithms to compute these paths. While the exposition uses closed polygonal regions $\Omega$, this restriction can be removed and we can simply treat the point-pairs of $\Omega$ as constraint edges. In fact, all our results can be restated for infinite lattice triangulations that differ by finitely many edges from the infinite equilateral lattice triangulation. 
    
    While out of the scope of this paper, significant consequences of our structural result are that any minimum flip plan generated by our algorithms (or simple modifications of them) is in fact \emph{unique}, as it the least-restrictive (minimum height) poset: i.e. not only are all of its consistent linear orderings shortest flip paths, by definition, but in fact every shortest flip path is consistent with the minimum flip plan.  
    Both the size and height of the least-restrictive minimum flip plans given in this paper are in fact metrics on the space of lattice triangulations (satisfy the triangle inequality).  
    This solves the optimal simultaneous flip path problems for lattice triangulations according to the measures studied in both the combinatorial \cite{galtier2003simultaneous,souvaine2011simultaneously} and geometric \cite{bose2007simultaneous,de2021transforming} settings.
    
    The motivation in \cite{caputo2015} for studying lattice triangulations is to prove bounds for the mixing times of Markov chains over weighted lattice triangulations.  
    This weighting determines whether the Markov chain favors transitions to lattice triangulations containing short edges or to those containing long edges. In both cases, mixing time bounds expressed in terms of the size of the point-set are conjectured, and weak versions of these conjectures are proved.  
    It is suggested that a deeper look into the structure of flip paths between lattice triangulations is necessary to prove the full conjecture, and we believe that the structural results about flip plans proved in this paper are sufficient for this purpose.
    
\section*{Acknowledgments}
    We thank Maxwell Nolan for computational experiments that led to some of the questions answered in this paper.

\bibliographystyle{elsarticle-num}
\bibliography{main}

\appendix

\section{Proof Details for Section \ref{section:single_edge}}
\label{appendix:sec4}

\subsection{Proof of Lemma \ref{lem:extend_from_pairs}}
\label{sec:proof_extend_from_pairs}

The proof of Lemma \ref{lem:extend_from_pairs} requires Lemmas \ref{lem:can_flip}, \ref{lem:length_increasing}, and \ref{lem:extend_to_maximal}, below.

    \begin{lemma}[Flippable Farey Parallelograms]
    \label{lem:can_flip}
    Consider a point-pair $(g,u)$ in an equilateral lattice point-set such that the Farey plan for $g$ has length at least $2$.  
    Let the longer point-pairs in the Farey parallelogram for $(g,u)$ be $(g_1,u)$ and $(g_1,v)$ and let $g_2=g-g_1$.  
    If some flip path $p$ between an equilateral triangulation and some triangulation $T'$ contains flips on the Farey parallelograms $P_{g_1,u}$ and $P_{g_1,v}$ and if $(g_1,u)$, $(g_2,u)$, $(g_1,v)$, $(g_2,u+g_1)$, and $(g_1-g_2,v)$ are edges in $T'$, then a flip on $P_{g,u}$ can be performed in $T'$. 
  \end{lemma}  
  
  \begin{proof}
    Since the Farey plan for $g$ has length at least $2$, the point-pairs in the Farey parallelogram $P_{g,u}$ are edges in $T'$, by Lemma \ref{lem:two_halves}.  
    Also, since $g$ does not belong to the equivalence classes $(0,1)$ or $(1,1)$, $g_1$ is not unit-length and the Farey parallelogram $P_{g_1}$ exists.  
    Consider the two triangulations $T_1$ and $T_2$ resulting from the flips in $p$ on the Farey parallelograms $P_{g_1,u}$ and $P_{g_1,v}$, respectively.  
    Since these flips were able to be performed, the regions in $T_1$ bounded by $P_{g_1,u}$ and $P_{g_1,v}$ contain only the edges $(g_1,u)$ and $(g_1,v)$, respectively.  
    Thus, by Lemma \ref{lem:two_halves}, the region in $T'$ bounded by $P_{g,u}$ contains only the shorter diagonal edge of $P_{g,u}$, and so a flip on $P_{g,u}$ can be performed in $T'$.
  \end{proof}
    
      \begin{lemma}[Length-increasing Flips]
  \label{lem:length_increasing}
  Consider the poset $\pi_{g,u}$ output by Algorithm Flip Plan on an input point-pair $(g,u)$.
  \begin{enumerate}
      \item All flips in $\pi_{g,u}$ replace edges with longer ones.
      \item All flips in $\pi_{g,u}$ that are not minimal add edges that are longer than the edges added by their child flips.
      \item The only flip in $\pi_{g,u}$ that adds an edge that is longer than a longer point-pair in the Farey parallelogram $P_{g,u}$ is the flip on $P_{g,u}$.
  \end{enumerate}
  \end{lemma}
  
  \begin{proof}
    Statement (1) is immediate from the construction of the poset $\pi_{g,u}$.  
    Statement (2) follows directly from repeated application of Lemma \ref{lem:flp_neighbor}.  
    Lastly, Statement (3) is a consequence of Statement (2).
  \end{proof}
    
    \begin{lemma}[Extending Adjacent Flip Plans]
    \label{lem:extend_to_maximal}
    Let $(g,u)$ and $(g,v)$ be adjacent point-pairs in an equilateral lattice point-set such that the Farey plan for $g$ has length at least $2$.  
      Also, let $(g_1,u_1)$ and $(g_1,u_2)$ be the longer point-pairs in the Farey parallelogram for $(g,u)$ and let $(g_1,u_3)$ and $(g_1,u_4)$ be the longer point-pairs in the Farey parallelogram for $(g,v)$.
      Lastly, on inputs $(g,u)$ and $(g,v)$, let $\pi_{g,u}$ and $\pi_{g,v}$ be the posets output by Algorithm Flip Plan, respectively.
      Then, Statement (1) implies Statement (2) below:
    \begin{enumerate}
          \item The poset $\pi_{g,u} \cup \pi_{g,v}$ with its two maximal flips removed is a flip plan for $(g_1,u_1)$, $(g_1,u_2)$, $(g_1,u_3)$, and $(g_1,u_4)$.
          \item The poset $\pi_{g,u} \cup \pi_{g,v}$ is a flip plan for $(g,u)$ and $(g,v)$.
      \end{enumerate}
    \end{lemma}

\begin{proof}
      Assume Statement (1) is true and consider any consistent linear ordering $p$ of $\pi_{g,u} \cup \pi_{g,v}$.  
      This sequence of flips can be split into subsequences $p=p_1p'p_2p''$ such that $p'$ and $p''$ are flips on the Farey parallelograms $P_{g,u}$ and $P_{g,v}$, respectively.  
      By assumption, $p_1$ is a flip path that starts from the equilateral triangulation $T$.  
      We will show that (i) $p_1p'$, (ii) $p_1p'p2$, and (iii) $p_1p'p_2p''$ are flip paths starting from $T$ one at a time.  
      Then, since the target triangulation of the flip path $p_1p'$ contains $(g,u)$ and no flip in $\pi_{g,u} \cup \pi_{g,v}$ removes $(g,u)$, by Lemma \ref{lem:length_increasing} (1) and (3), the target triangulation of $p$ contains $(g,u)$ and $(g,v)$.  
      Thus, the lemma is proved.
      
      For (i), observe that $p_1$ contains all flips in the poset $\pi_{g,u}$, except for the maximal flip on $P_{g,u}$.  
      Since the Farey plan for $g$ has length at least $2$, the Farey parallelogram for $g_1$ exists.
      Hence, by Lemma \ref{lem:flp_neighbor}, $p_1$ contains a flip $s$ on $P_{g_1,u_1}$, and the triangulation resulting from $s$ contains $(g_1,u_1)$ and the point-pairs in $P_{g_1,u_1}$ as edges.  
      If some flip in $p_1$ after $s$ removes one of these edges, then the Farey parallelogram of the first flip that does so contains $(g_1,u_1)$, by Lemma \ref{lem:length_increasing} (1).  
      The edge added by this flip is longer than $(g_1,u_1)$, so, by Lemma \ref{lem:length_increasing} (3), this flip must be $p'$, which is a contradiction.  
      Therefore, no flip in $p_1$ after $s$ removes $(g_1,u_1)$ or an edge in $P_{g_1,u_1}$
      This implies that the target triangulation $T'$ of $p_1$ contains $(g_1,u_1)$ and the point-pairs in $P_{g_1,u_1}$ as edges.  
      Similarly, $T'$ contains $(g_1,u_2)$ and the point-pairs in $P_{g_1,u_2}$ as edges.  
      Consequently, by Lemma \ref{lem:can_flip}, the flip $p'$ can be performed after $p_1$, and so (i) is true.
      
      For (ii), if $p_2$ is empty, then (ii) follows from (i).  
      Otherwise, assume that some flip along $p_1p'p_2$ cannot be performed, and let $s$ be the first such flip, which adds the edge $(h,y)$.  
      By (i), $s$ is contained in both $p_2$ and $\pi_{g_1,v}$.  
      Additionally, by assumption, $s$ can be performed along the flip path $p_1p_2$.  
      Hence, the point-pairs in the Farey parallelogram $P_{h,y}$ are edges of the triangulation preceding $s$ in $p_1p_2$.  
      The facts above imply that the flip $p'$ on $P_{g,u}$ removes one of these edges, which must be the shorter diagonal of $P_{g,u}$.  
      However, as argued above, $T'$ contains $(g_1,u_1)$, $(g_1,u_2)$, $P_{g_1,u_1}$, and $P_{g_1,u_2}$ as edges.  
      A similar argument shows that no flip in $p_1p_2$ removes any of these edges, so they are edges in the triangulation preceding $s$ along $p_1p_2$.
      Therefore, $P_{h,y}$ contains one of the longer point-pairs in $P_{g,u}$.  
      This implies that $s$ adds an edge whose vector is longer than the vector of this point-pair, and so $s$ is the flip $p''$, by Lemma \ref{lem:length_increasing} (3).  
      This is a contradiction, so (ii) is true.  
      
      For (iii), we can use (ii) and an argument similar to (ii) to show that the point-pairs in $P_{g,v}$ and the shorter diagonal of $P_{g,v}$ are edges in the target triangulation of the flip path $p_1p'p_2$ and no flip in this flip path removes an edge in $P_{g,v}$ or the shorter diagonal of $P_{g,v}$.  
      Thus, (iii) follows from Lemma \ref{lem:can_flip}, and the lemma is proved.  
    \end{proof}
    
    \begin{proof}[Proof of Lemma \ref{lem:extend_from_pairs}]
      It suffices to show that Statements (1) and (2) imply Statement (1) of Lemma \ref{lem:extend_to_maximal}, and then the lemma follows by applying By Lemma \ref{lem:extend_to_maximal}.  
      Let $\pi'$ be the poset $\pi_{g,u} \cup \pi_{g,v}$ with its maximal two flips removed and let $p=p_1,\dots,p_n$ be any consistent linear ordering of $\pi'$.  
      Assume that, starting from the equilateral triangulation $T$, $p_t$ is the first flip in $p$ that cannot be performed, and let $(h,y)$ be the edge added by this flip.  
      By assumption, the subsequence $p'=p_1,\dots,p_{t-1}$ is a flip path that starts from $T$.  
      Note that if $t = n$, then we can use Lemma \ref{lem:length_increasing} (1) and (3) to show that Statement (1) of Lemma \ref{lem:extend_to_maximal} is true, and so we are done.  
      Hence, assume that $t < n$.  
      
      Next, if $p_t$ is a minimal flip in $\pi'$, then $(h,y)$ belongs to the equivalence class $(1,1)$ and its Farey parallelogram $P_{h,y}$ consists of unit-length point-pairs.  
      Observe that the point-pairs in $P_{h,y}$ are edges of $T$, and the region bounded by these edges contains only a unit-length edge.  
      Hence, some flip in $p'$ must remove an edge in $P_{h,y}$.  
      The first flip that does so adds an edge whose Farey parallelogram contains some edge in $P_{h,y}$ and the shorter diagonal of $P_{h,y}$.  
      This implies that the defining coordinate pair of the added edge is different than that of $(h,y)$.  
      However, by construction, all edges added by flips in $\pi'$ have the same defining coordinate pair, so we get a contradiction.  
      
      Therefore, assume that $p_t$ is not a minimal flip in $\pi'$.
      Also, let $P_{h,y}=\{(h_1,y),(h_2,y+h_1),(h_2,y),(h_1,y+h_2)\}$, where $h_1$ is longer than $h_2$.  
      Since $p_t$ is not a minimal flip, the Farey plan for $h$ has length at least $2$, and so the Farey parallelograms $P_{h_1,y}$ and $P_{h_1,y+h_2}$ exist.  
      Hence, the shorter diagonal of $P_{h,y}$ is $(h_1-h_2,y+h_2)$, and it is shared by $P_{h_1,y}$ and $P_{h_1,y+h_2}$, by Lemma \ref{lem:two_halves}.  
      By Lemma \ref{lem:flp_neighbor}, $p'$ contains a flip $p_s$ on $P_{h_1,y}$, and the triangulation resulting from $p_s$ contains $(h_1,y)$, $(h_2,y)$, and $(h_1-h_2,y+h_2)$ as edges.  
      Similarly, some triangulation resulting from another flip in $p'$ contains $(h_1,y+h_2)$, $(h_2,y+h_1)$, and $(h_1-h_2,y+h_2)$ as edges.  
      
      Finally, if the target triangulation $T'$ of $p'$ contains the point-pairs in $P_{h,y}$ and the shorter diagonal of $P_{h,y}$ as edges, then $p'p_t$ is a flip path that starts from $T$, by Lemma \ref{lem:can_flip}, and we are done.  
      Otherwise, some flip $p_r$ in $p'$, with $s+1 \leq r \leq t-1$, removes either $(h_1,y)$, $(h_2,y)$, or $(h_1-h_2,y+h_2)$.  
      Consider the case where some flip plan $\pi_i$ contains both $p_r$ and $p_t$, for some integer $1 \leq i \leq 4$.  
      Then, the maximal subsequence $p''p_t$ of $p'p_t$ containing only flips in $\pi_i$ is a flip path that starts from $T$, by assumption.  
      However, since $p''$ contains $p_r$, Lemma \ref{lem:length_increasing} (1) tells us that the target triangulation of $p''$ does not contain either some point-pair in $P_{h,y}$ or the shorter diagonal of $P_{h,y}$ as an edge.  
      This implies that $p''p_t$ is not a flip path, which is a contradiction.  
      We arrive at a similar contradiction if some flip plan $\pi_i \cup \pi_j$ contains both $p_r$ and $p_t$, for some distinct integers $1 \leq i,j \leq 4$.  
      Since these cases are exhaustive, the lemma is proved.  
    \end{proof}
    
    \subsection{Proof of Lemma \ref{lem:bounding_parallelogram}}
    \label{sec:proof_bounding_parallelogram}
    
    \begin{proof}[Proof of Lemma \ref{lem:bounding_parallelogram}]
  We proceed by induction on the size $n$ of the Farey plan for the vector $g$.  
  Let $\pi_{g,u}$ be the poset output by Algorithm Flip Plan on input $(g,u)$.  
  Since $(g,u)$ is not unit-length, $n$ is at least $1$ and $\pi_{g,u}$ is non-empty.  
  When $n=1$, $(g,u)$ belongs to the equivalence class $(1,1)$ and its bounding region is its Farey parallelogram.  
  The only flip in $\pi_{g,u}$ is on this parallelogram, so the base case holds.  
  
  Next, assume the lemma holds when $n = k$, for any $k \geq 1$, and we will show it holds for $n=k+1$.  
  Clearly, the Farey parallelogram of the maximal flip in $\pi_{g,u}$ is contained in the bounding region for $(g,u)$.  
  Consider the longer point-pairs $(g_1,u)$ and $(g_1,v)$ in the Farey parallelogram $P_{g,u}$.  
  By Lemma \ref{lem:farey_plan_longer_vector}, the Farey plan for $g_1$ has size $k$.  
  Therefore, by Lemma \ref{lem:flp_neighbor} and the inductive hypothesis, the bounding regions for $(g_1,u)$ and $(g_1,v)$ contain all Farey parallelograms of the flips in $\pi_{g,u}$, except for the Farey parallelogram $P_{g,u}$ of the maximal flip.  
  Lastly, by the definition of the inverse Farey-Flip map, $(g_1,u)$ and $(g_1,v)$ have the same defining coordinate pair as $(g,u)$, so their bounding regions are clearly contained in the bounding region for $(g,u)$.  
  Thus, the lemma is proved by induction.
  \end{proof}
    
\section{Proof Details for Section \ref{section:multi_edge}}
\label{appendix:sec_5}

\subsection{Proof of Lemma \ref{lem:farey_parallelogram_intersect}}
\label{sec:proof_farey_parallelogram_intersect}

\begin{proof}[Proof of Lemma \ref{lem:farey_parallelogram_intersect}]
  We proceed by induction on the length of the Farey plan for the vector $g$, say $n$.  
  Since $g$ is not unit-length, $n$ is at least $1$.  
  When $n = 1$, $(g,u)$ belongs to the equivalence class $(1,1)$ and the flip plan $\pi_{g,u}$ contains a single flip on the Farey parallelogram $P_{g,u}$.  
  Clearly, the line-segment of $(g,u)$ intersects the line-segment of the shorter diagonal of $P_{g,u}$.  
  
  Next, assume the lemma is true when $n=k$, for any $k \geq 1$, and we will show it is true when $n=k+1$.  
  The maximal flip in $\pi_{g,u}$ is on $P_{g,u}$, so the line-segment of $(g,u)$ intersects the line-segment of the shorter diagonal of $P_{g,u}$.  
  Let $(g_1,u)$ and $(g_1,v)$ be the longer point-pairs in $P_{g,u}$ and consider the flip plans $\pi_{g_1,u}$ and $\pi_{g_1,v}$.  
  By Lemma \ref{lem:farey_plan_longer_vector}, the inductive hypothesis states that the line-segment of the shorter diagonal of every Farey parallelogram in $\pi_{g_1,u}$ and $\pi_{g_1,u}$ intersects the line-segments of $(g_1,u)$ and $(g_1,v)$, respectively.  
  Wlog, if one of the line-segments of the shorter diagonal of some Farey parallelogram $P$ in $\pi_{g_1,u}$ does not intersect the line-segment of $(g,u)$, then either $P$ is not contained the the bounding region for $(g_1,u)$ or we can use Lemma \ref{lem:two_halves} to show that the region bounded by $P_{g,u}$ contains a lattice point.  
  The former case contradicts Lemma \ref{lem:bounding_parallelogram} and the latter case contradicts Corollary \ref{cor:unique_quad}.  
  Therefore, the shorter diagonal of every Farey parallelogram in $\pi_{g_1,u}$ and $\pi_{g_1,u}$ intersects the line-segments of $(g,u)$.  
  Finally, by Lemma \ref{lem:flp_neighbor}, $\pi_{g_1,u} \cup \pi_{g_1,v}$ is $\pi_{g,u}$ with its maximal flip removed, and so the lemma is proved by induction.  
\end{proof}

\subsection{Proof of Lemma \ref{lem:flip_plan_for_independent}}
\label{sec:proof_flip_plan_for_independent}

The proof of Lemma \ref{lem:flip_plan_for_independent} requires a theorem given in \cite{dyn1993transforming} and Lemmas \ref{lem:farey_parallelogram_intersect}, \ref{lem:multi_flip_minimum}, and \ref{lem:important_path}, below.  
The idea is to show that if two point-pairs $(g,u)$ and $(g,v)$ do not have intersecting line-segments, then the line segment of $(g,u)$ does not intersect the line-segment of a point-pair added by a flip in the flip plan $\pi_{g,v}$.  

\begin{theorem}[\cite{dyn1993transforming}, Theorem 2.1]
\label{thm:dyn}
Let $(g,u)$ be a point-pair that is an edge in some triangulation of in a lattice point-set $S$.  
Then, for any triangulation $T$ of $S$, there exists a flip path $p$ that starts from $T$ and forces $(g,u)$ to become an edge such that all edges removed by flips in $p$ intersect the line-segment of $(g,u)$.
\end{theorem}

\begin{lemma}[The Farey Parallelograms Resulting from Algorithm Multi Flip Plan]
\label{lem:multi_flip_minimum}
    Let $G$ be a set of point-pairs that are edges in some triangulation of an equilateral lattice point-set and let $G'$ be its maximum independent subset.  
    Also, let $\pi_G$ be the poset output by Algorithm Multi Flip Plan on input $G$.  
    If some consistent linear ordering $p$ of $\pi_G$ is a flip path for $G$, then $G'$ is the set of all non-unit-length edges of the target triangulation $T'$ of $p$ such that the point-pairs in each of their Farey parallelograms are edges of $T'$.  
    Furthermore, if $G'$ is empty, then every edge of $T'$ is unit-length.  
\end{lemma}

\begin{proof}
    If $G'$ is empty, then $\pi_G$ is empty.  
    This implies that $T' = T$ is the equilateral triangtulation.  
    Hence, by definition, every edge of $T'$ is unit-length, and so the lemma is proved.  

    Therefore, assume $G'$ is non-empty.  
   Then, the maximal flips in $\pi_G$ are on the Farey parallelograms for point-pairs in $G'$, by definition.  
   Hence, for any point-pair $(g_i,u_i)$ in $G'$, the triangulation resulting from the maximal flip on the Farey parallelogram $P_{g_i,u_i}$ contains the point-pairs in $P_{g_i,u_i}$ as edges.  
   If $T'$ does not contain the point-pairs in $P_{g_i,u_i}$ as edges, then some flip in $\pi_G$ removes an edge in $P_{g_i,u_i}$.  
   By Lemma \ref{lem:length_increasing} (1), the first flip to do so is performed on a Farey parallelogram containing $(g_i,u_i)$ as a longer point-pair.  
   Consequently, Lemma \ref{lem:flp_neighbor} implies that the flip on $P_{g_i,u_i}$ is not maximal in $\pi_G$, which is a contradiction.  
   Thus, $T'$ contains the point-pairs in the Farey parallelograms for the point-pairs in $G'$ as edges.  
   
   To see that these are the only non-unit-length edges in $T'$ such that the point-pairs in their Farey parallelograms are edges of $T'$, first recall that every edge of $T$ is unit-length.  
   Hence, the only candidate edges are those added by flips in $\pi_G$, other than the ones added by the maximal flips.  
   However, using Lemma \ref{lem:farey_parallelogram_intersect}, we see that some edge in $G'$ intersects the line-segment of some point-pair in the Farey parallelogram for each of these edges.  
   Thus, for each of these edges, some point-pair in its Farey parallelogram is not an edge of $T'$, and so the lemma is proved.
\end{proof}

\begin{lemma}[Extending Flip Paths for Point-pairs]
\label{lem:important_path}
    Let $G=\{(g_1,u_1),\dots,$ \\ $(g_{n+1},u_{n+1})\}$ be an independent set of point-pairs that are edges in some triangulation of a lattice point-set and let $\pi_{G_n}$ be the poset output by Algorithm Multi Flip Plan on input $G_n = G \setminus \{(g_{n+1},u_{n+1})\}$.  
    Also, let some consistent linear ordering of $\pi_{G_n}$ be a flip path between an equilateral triangulation and a triangulation $T'$ that contains $G_n$, but not $(g_{n+1},u_{n+1})$, as edges.  
    Then, there exists a flip path $p$ that starts from $T'$ and forces $(g_{n+1},u_{n+1})$ to become an edge such that all edges replaced by flips in $p$ intersect the line-segment of $(g_{n+1},u_{n+1})$ and $p$ contains all flips in $\pi_{g_{n+1},u_{n+1}} \setminus \pi_{G_n}$.
\end{lemma}

\begin{proof}
   Let $\pi' = \pi_{g_{n+1},u_{n+1}} \setminus \pi_{G_n}$.  
   Since $G$ is independent, $(g_{n+1},u_{n+1})$ is not unit-length, and so $\pi'$ is non-empty.  
   Also, since $G$ are edges in some triangulation, Theorem \ref{thm:dyn} tells us that there exists a flip path $p=p_1,\dots,p_m$ that starts from $T'$ and forces $(g_{n+1},u_{n+1})$ to become an edge such that all edges replaced by flips in $p$ intersect the line-segment of $(g_{n+1},u_{n+1})$.  
   Note that $p$ is non-empty, since $(g_{n+1},u_{n+1})$ is not an edge of $T'$.  
   
   First, we show that we can remove flips from $p$ to obtain a sequence $p'$ such that all flips in $p'$ replace edges with longer ones and $p'$ is a flip path that starts from $T'$ and forces $(g_{n+1},u_{n+1})$ to become an edge.  
   Assume that $p_t$ is the first flip in $p$ that replaces an edge $(h,y)$ with a shorter one.  
   By Corollary \ref{cor:unique_quad}, $p_t$ is a flip on a Farey parallelogram.  
   Using Lemma \ref{lem:multi_flip_minimum}, we see that $G_n$ is the set of all edges of $T'$ such that the point-pairs in their Farey parallelograms are edges of $T'$.  
   Also, since all edges replaced by flips in $p$ intersect the line-segment of $(g_{n+1},u_{n+1})$, no point-pair in $G_n$ is removed by a flip in $p$, by our assumptions.  
   The facts above imply that $t>1$ and some flip $p_s$ in $p$, with $1 \leq s < t$, adds the edge $(h,y)$ that is removed by $p_t$.  
   
   Next, if $(h,y)$ is $(g_{n+1},u_{n+1})$, then $p'=p_1,\dots,p_s$ is the desired flip path.  
   Otherwise, observe that the triangulations resulting from $p_s$ and $p_t$ contain the point-pairs in $P_{h,y}$ as edges.  
   Furthermore, since all flips between $p_s$ and $p_t$ replace edges with longer ones, $(h,y)$ and the point-pairs in $P_{h,y}$ must be edges in the triangulations resulting from these flips.  
   Thus, removing $p_s$ and $p_t$ from $p$ yields a flip path between the same triangulations as $p$ and containing $1$ less flip that replaces an edge with a shorter one.  
   We can repeat this process to obtain the desired flip path $p'$.  
   
   Finally, we show that $p'$ contains the flips in $\pi'$.  
   Using Lemma \ref{lem:length_increasing} (1) and Corollary \ref{cor:unique_quad}, we see that $T'$ does not contain any edge added by a flip in $\pi'$.  
   Since the flips in $p'$ replace edges with longer ones, Corollary \ref{cor:unique_quad} tells us that $p'$ contains a flip on the Farey parallelogram $P_{g_{n+1},u_{n+1}}$.  
   Note that this is the maximal flip in $\pi'$.  
   Next, let $\pi'_1$ be $\pi'$ with this flip removed.  
   If $\pi'_1$ is non-empty, then the triangulation $T'$ does not contain some longer point-pair in $P_{g_{n+1},u_{n+1}}$ as an edge, by Lemma \ref{lem:flp_neighbor}.  
   Hence, $p'$ must contain a flip on the Farey parallelogram for this point-pair, which is a maximal flip in $\pi'_1$.  
   By repeating this argument, it is clear that $p'$ contains all flips in $\pi'$.
\end{proof}

\begin{proof}[Proof of Lemma \ref{lem:flip_plan_for_independent}]
  We proceed by induction on $n=|G|$.  
  When $n=1$, if $G'$ is empty, then $\tau_G$ is empty, and the lemma is immediate.  
  Otherwise, we have $G'=\{(g_1,u_1)\}$ and $\tau_G = \pi_{g_1,u_1}$, and so the lemma holds by Corollary \ref{cor:EdgeCreate}.  
  Next, for any integers $k \geq 1$ and $1 \leq j \leq k$, assume that the lemma holds when $n=j$.  
  We will prove the lemma when $n=k+1$.  
  Consider the set $G_k = \{(g_1,u_1),\dots,(g_k,u_k)\}$, the poset $\tau_{G_k}=\pi_{g_1,u_1}||\dots||\pi_{g_k,u_k}$, and the maximum independent subset $G'_k$ of $G_k$.  
  By the inductive hypothesis, $\tau_{G_k}$ is a flip plan for $G'_k$.  

  There are two cases.  
  
  \medskip
  \noindent\textbf{Case 1:} $(g_{k+1},u_{k+1})$ is not contained in $G'$.
  \medskip
  
  In this case, we have $G'_k = G'$.  
  Along with the definition of an independent set, this implies that $\tau_{G_k}=\tau_G$.
  Hence, the lemma follows from the fact that $\tau_{G_k}$ is a flip plan for $G'_k$.
  
  \medskip
  \noindent\textbf{Case 2:}  $(g_{k+1},u_{k+1})$ is contained in $G'$.
  \medskip
  
  In this case, $(g_{k+1},u_{k+1})$ is not unit-length, by the definition of $G'$.  
  Consider the set $H$ of point-pairs $(g_i,u_i) \in G$ such that $\pi_{g_i,u_i}$ is a subposet of $\pi_{g_{k+1},u_{k+1}}$ but not a subposet of $\pi_{g_j,u_j}$, for any $(g_j,u_j) \in G_k \setminus \{(g_i,u_i)\}$.  
  Then, we have $G'_k = (G' \cup H) \setminus \{(g_{k+1},u_{k+1})\}$.  
  Observe that each point-pair in $H$ is added as an edge by some flip in $\pi_{g_{k+1},u_{k+1}}$, and no two of these flips are comparable in this flip plan.  
  This implies that, for any two point-pairs in $H$, some consistent linear ordering of $\pi_{g_{k+1},u_{k+1}}$ contains two consecutive flips that add two of these point-pairs as edges, and the triangulation resulting from the second flip contains both point-pairs as edges.  
  Hence, no two point-pairs in $H$ have intersecting line-segments, and so $H$ are edges of some triangulation.  
  
  Next, if $I = G' \setminus \{(g_{k+1},u_{k+1})\} = \{(g'_1,u'_1),\dots,(g'_{\ell},u'_{\ell})\}$ is empty, then $G'=\{(g_{k+1},u_{k+1})\}$ and $\tau_{G} = \pi_{g_{k+1},u_{k+1}}$, and so the lemma follows from Corollary \ref{cor:EdgeCreate}.  
  Otherwise, since no two point-pairs in $I$ have intersecting line-segments, by assumption, $I$ is a subset of edges of some triangulation.  
  Hence, applying the inductive hypothesis to $I$ shows that $\tau_I = \pi_{g'_1,u'_1} || \dots || \pi_{g'_{\ell},u'_{\ell}}$ is a flip plan that starts from $T$ and forces $I$ to become edges.  
  Observe that (i) $I \cup \{(g_{k+1},u_{k+1})\}$ is independent and is a subset of edges of some triangulation, (ii) any consistent linear ordering of $\tau_I$ is a consistent linear ordering of the poset output by Algorithm Multi Flip Plan on input $I$, and (iii) the target triangulation $T'$ of $\tau_I$ contains $I$, but not $(g_{k+1},u_{k+1})$, as edges.  
  Therefore, by Lemma \ref{lem:important_path}, there exists a flip path $r$ that starts from $T'$ and forces $(g_{k+1},u_{k+1})$ to become an edge.  
  Moreover, $r$ contains the flips in $\pi_{g_{k+1},u_{k+1}} \setminus \tau_I$ and all edges removed by flips in $r$ intersect the line-segment of $(g_{k+1},u_{k+1})$.  
  
  Finally, since $I \cup \{(g_{k+1},u_{k+1})\}$ is an independent set of point-pairs, no point-pair in $I$ is removed by a flip in $r$.  
  Hence, since each point-pair in $H$ is added as an edge by a flip in $r$, the line-segment of this point-pair does not intersect the line-segment of any point-pair in $I$.  
  This implies that $G'_k = I \cup H$ is a subset of edges of some triangulation.  
  Therefore, we can apply the inductive hypothesis to $G'_k$ to get that $\tau_{G_k}$ is a flip plan that starts from $T$ and forces $G'_k$ to become edges.  
  Let $T''$ be the target triangulation of $\tau_{G_k}$.  
  By the definition of poset concatenation, we must show that $\pi' = \pi_{g_{k+1},u_{k+1}} \setminus \tau_{G_k}$ is a flip plan that starts from $T''$ and forces $(g_{k+1},u_{k+1})$ to become an edge such that the target triangulation contains $G'$ as edges.  
  Observe that no flip in $\pi'$ removes a point-pair in $G'$, or else the line-segments of $(g_{k+1},u_{k+1})$ and this point-pair intersect, by Lemma \ref{lem:farey_parallelogram_intersect}, contradicting our assumption.  
  Consequently it suffices to show that $\pi'$ is a flip plan that starts from $T''$ and forces $(g_{k+1},u_{k+1})$ to become an edge.  

  Assume to the contrary that $\pi'$ is not a flip plan that starts from $T''$ and forces $(g_{k+1},u_{k+1})$ to become an edge.  
  Then, some consistent linear ordering $p=p_1,\dots,p_m$ of $\pi'$ is not a flip path that starts from $T''$ and forces $(g_{k+1},u_{k+1})$ to become an edge.  
  Let $p_t$ is the first flip in $p$ that cannot be performed, and let $(h,y)$ be the edge that it adds.  
  Note that since $\tau_{G_k}$ is a flip plan that starts from $T$ and forces $G'_k$ to become edges, $p' = p_1,\dots,p_{t-1}$ is a flip path that starts from $T''$ and forces $H$ to become edges.  
  There are two subcases.  
  
  \medskip
  \noindent\textbf{Subcase 1:} $p_t$ is not a minimal flip in $\pi'$ and its child flips in $\pi_{g_{k+1},u_{k+1}}$ are contained in $\pi'$.
  \medskip
  
  By assumption, $p'$ contains the child flips of $p_t$.  
  Using Lemmas \ref{lem:two_halves} and \ref{lem:flp_neighbor}, we see that the triangulations resulting from each child flip contains half of the point-pairs in the Farey parallelogram $P_{h,y}$ as edges.  
  If any flip in $p'$ removes any one of these edges, then the target triangulation of $p'$ does not contain some point-pair in $P_{h,y}$ as an edge, by Lemma \ref{lem:length_increasing} (1).  
  However, since all flips in $p'$ are contained in $\pi'$, this implies that $\pi_{g_{k+1},u_{k+1}}$ is not a flip plan, contradicting Corollary \ref{cor:EdgeCreate}.  
  
  \medskip
  \noindent\textbf{Subcase 2:} $p_t$ is either a minimal flip in $\pi'$ or at least one of its child flips in $\pi_{g_{k+1},u_{k+1}}$ is not contained in $\pi'$.  
  \medskip
  
  Let $q$ be any consistent linear ordering of $\tau_{G_k}$.  
  If $p_t$ is a minimal flip in $\pi_{g_{k+1},u_{k+1}}$, then $P_{h,y}$ and its shorter diagonal are unit-length point-pairs that are edges in $T$.  
  Otherwise, the child flips of $p_t$ are contained in $qp'$, and the triangulations resulting from these flips each contain half of the point-pairs in $P_{h,y}$ as edges, by Lemmas \ref{lem:two_halves} and \ref{lem:flp_neighbor}.
  Hence, in either case, each point-pair in $P_{h,y}$ is an edge in some triangulation along the flip path $qp'$.  
  If the target triangulation of $qp'$ contains the point-pairs in $P_{h,y}$ as edges, then $p'p_t$ is a flip path starting from $T''$.  
  Otherwise, some flip $q_s$ in $qp'$ is the first to remove one of these edges, say $(h_1,y_1)$.  
  
  Next, if $q_s$ is contained in $\pi_{g_{k+1},u_{k+1}}$, then we arrive at a contradiction similar to the one in Subcase 1.  
  Hence, $q_s$ is contained in $\pi_{g'_i,u'_i} \setminus \pi_{g_{k+1},u_{k+1}}$, for some $(g'_i,u'_i) \in I$.  
  By Lemma \ref{lem:farey_parallelogram_intersect}, the line-segments of $(g'_i,u'_i)$ and $(h_1,y_1)$ intersect.  
  This implies that any flip path that starts from a triangulation containing $(g'_i,u'_i)$ as an edge and contains a flip is on $P_{h,y}$ must contain a flip that removes $(g'_i,u'_i)$.  
  
  Finally, recall from the discussion above that the triangulation $T'$ contains $I$, and hence $(g'_i,u'_i)$, as edges, and the flip path $r$ starts from $T'$ and forces $(g_{k+1},u_{k+1})$ to become an edge.  
  Additionally, $r$ contains the flips in $\pi'$, which includes a flip on $P_{h,y}$.  
  Hence, by the argument above, $r$ contains a flip that removes $(g'_i,u'_i)$.  
  However, by construction, all edges removed by flips in $r$ intersect the line-segment of $(g_{k+1},u_{k+1})$.  
  Thus, the line-segments $(g'_i,u'_i)$ and $(g_{k+1},u_{k+1})$ intersect, which contradicts our assumption.  
  
  \medskip
  The subcases above are exhaustive, and the contradictions imply that $\pi'$ is a flip plan that starts from $T''$ and forces $(g_{k+1},u_{k+1})$ to become an edge.  
  This completes Case 2.  
  Thus, since the above cases are exhaustive, the lemma is proved by induction.  
\end{proof}

\section{Proof Details for Section \ref{sec:ground_state}}
\label{appendix:sec6}

\subsection{Proof of Lemma \ref{lem:unique_minimum_tri}}
\label{sec:proof_unique_minimum_tri}

We require Lemmas \ref{lem:flp_target_minimum} and
\ref{lem:bijection_tri_minimum} to prove Lemma  \ref{lem:unique_minimum_tri}.  
The following lemmas are used to prove these lemmas.  
Recall the definition of the maximum independent set of a triangulation, given in Section \ref{sec:ground_state}.

\begin{lemma}[Target Triangulation of Flip Plans]
\label{lem:flp_target}
    If $T$ is a triangulation of an equilateral lattice point-set whose maximum independent set is $G$, then $T$ is the target triangulation of the flip plan $\pi_G$.
\end{lemma}

\begin{proof}
   Consider the set $G'$ of all edges in $T$ and the flip plan $\pi_{G'}$.
   By Theorem \ref{thm:MultiEdgeCreate}, $T$ is the target triangulation of $\pi_{G'}$.  
   The lemma follows by observing that $\pi_G = \pi_{G'}$.
\end{proof}

\begin{lemma}[Bijection Between Triangulations and Independent Sets of Point-pairs]
\label{lem:bijection_tri_ind_set}
    The map that takes a triangulation of an equilateral lattice point-set $S$ to its maximum independent set is a bijection between triangulations of $S$ and the union of the empty-set and all sets of point-pairs that are (i) independent in $S$ and (ii) edges in some triangulation of $S$.
\end{lemma}

\begin{proof}
    First, note that the map is well-defined.  
   Next, for injectivity, assume that two triangulations of $S$ are mapped to the same set of point-pairs.  
   Then, Lemma \ref{lem:flp_target} shows that these triangulations are the same.
   Finally, for surjectivity, consider any (possibly empty) set $G$ of point-pairs satisfying (i) and (ii).  
   Theorem \ref{thm:MultiEdgeCreate} and Lemma \ref{lem:multi_flip_minimum}, in \ref{sec:proof_flip_plan_for_independent}, show that $G$ is the maximum independent set of the minimum triangulation containing $G$ as edges. 
\end{proof}

\begin{lemma}[Target Triangulations of Flip Plans are Minimum Triangulations]
\label{lem:flp_target_minimum}
    If $G$ is a set of point-pairs that are edges in some triangulation of an equilateral lattice point-set, then the target triangulation of the flip plan $\pi_G$ is the unique minimum triangulation containing $G$ as edges.  
\end{lemma}

\begin{proof}
   By Lemma \ref{lem:multi_flip_minimum}, in \ref{sec:proof_flip_plan_for_independent}, the target triangulation $T$ of $\pi_G$ is a minimum triangulation containing $G$ as edges.  
   Consider any other triangulation $T'$ containing $G$ as edges.  
   By Lemma \ref{lem:bijection_tri_ind_set}, $T$ and $T'$ have distinct maximum independent sets $H$ and $H'$, respectively.  
   Also, by Lemma \ref{lem:flp_target}, $T'$ is the target triangulation of flip plan $\pi_{H'}$ and, by Lemma \ref{lem:multi_flip_minimum}, $T'$ contains an edge in $H' \setminus H$ that is contained in its Farey parallelogram.
   Thus, $T'$ is not a minimum triangulation containing $G$ as edges.
\end{proof}

\begin{lemma}[Identity Map Between Triangulations and Minimum Triangulations of an Equilateral Point-set]
\label{lem:bijection_tri_minimum}
    If $T$ is a triangulation of an equilateral lattice point-set $S$ whose maximum independent set is $G$, then the map that takes $T$ to the minimum triangulation of $S$ containing $G$ as edges is the identity map.  
\end{lemma}

\begin{proof}
    It suffices to show that the composition of the map that sends $T$ to $G$ and the map that sends $G$ to the minimum triangulation $MT(S,G)$ is the identity map.  
    First, observe that these maps are well-defined.  
    Second, by Lemmas \ref{lem:flp_target} and \ref{lem:flp_target_minimum}, the second map sends $G$ to $T$.  
\end{proof}

\begin{proof}[Proof of Lemma \ref{lem:unique_minimum_tri}]
   Let the lattice point-set $S=L \cap \Omega$ and let $G$ be a set of point-pairs that are edges in some triangulation of $S$.  
   If $S$ admits an equilateral triangulation, then the lemma follows immediately from Lemma \ref{lem:flp_target_minimum}.  
   Otherwise, consider a lattice point-set $S'$ containing $S$ that admits an equilateral triangulation.  
   Also, consider the set $\Omega \cup G$ and let $G'$ be its maximum subset of point-pairs that is independent in $S'$.  
   Observe that $G' \setminus \Omega$ is the maximum subset of point-pairs in $G$ that is independent in $S$, by definition.  
   
   Since $G$ is a subset of edges of some triangulation of $S$, and all triangulations of $S$ contain $\Omega$ as edges, $\Omega \cup G$ is a subset of edges of some triangulation of $S'$.  
   Hence, by Lemma \ref{lem:flp_target_minimum}, $\pi_{\Omega} \cup \pi_G$ is a flip plan between the equilateral triangulation of $S'$ and the minimum triangulation $MT(S',\Omega \cup G)$.  
   Also, by definition, the point-pairs in $G'$ are the only edges of $MT(S',\Omega \cup G)$ such that the point-pairs in their Farey parallelograms are edges of $MT(S',\Omega \cup G)$.  
   
   Next, consider the triangulation $T$ of $S$ obtained by deleting the lattice points in $S' \setminus S$ from $MT(S',\Omega \cup G)$.  
   Observe that the Farey parallelogram for each edge of $T$ in $\Omega$ contains a vertex in $S' \setminus S$, and so some point-pair in this parallelogram is not an edge of $T$.  
   On the other hand, consider the Farey parallelogram $P_{g,u}$ for any edge $(g,u)$ of $T$ in $G' \setminus \Omega$.  
   If $P_{g,u}$ is not contained in $S$, then the line segment of one of its point-pairs intersects the line-segment of a point-pair in $\Omega$.  
   This implies that some point-pair in $P_{g,u}$ is not an edge of $MT(S',\Omega \cup G)$, which is a contradiction.  
   Hence, since each point-pair in $P_{g,u}$ is an edge of $MT(S', \Omega \cup G)$, it is an edge of $T$.  
   Lastly, since some point-pair in the Farey parallelogram for any edge of $MT(S', \Omega \cup G)$ that is not contained in $G'$ is not an edge of $MT(S', \Omega \cup G)$, by definition, it is also not an edge of $T$.  
   Therefore, $G' \setminus \Omega$ is the set of all edges of $T$ such that the point-pairs in their Farey parallelograms are edges of $T$.  
   By definition, this implies that $T$ is a minimum triangulation $MT(S,G)$.  
   
   Finally, we show that $MT(S,G)$ is unique. 
   Let $T'$ be any other triangulation of $S$ containing $G$ as edges.  
   Also, consider the triangulation $T''$ of $S'$ obtained from $T'$ by adding the points and edges that we removed from $MT(S',\Omega \cup G)$, above.  
   By Lemma \ref{lem:bijection_tri_minimum}, $T''$ is the minimum triangulation for its maximum independent set, which is distinct from $G'$, since $T''$ is not $MT(S', \Omega \cup G)$.  
   Since $T''$ and $MT(S',\Omega \cup G)$ differ only in their edges between points in $S$, this implies that $T''$ contains some edge between points in $S$ that is not contained in $G' \setminus \Omega$ and such that the point-pairs in its Farey parallelogram are edges of $T''$.  
   Thus, this is also true of $T'$, and so $T'$ is not a minimum triangulation of $S$ containing $G$ as edges.
\end{proof}

\subsection{Proof of Lemma \ref{lem:constrained_flp_equilateral}}
\label{sec:proof_constrained_flp_equilateral}

\begin{proof}[Proof of Lemma \ref{lem:constrained_flp_equilateral}]
   First, we prove Statement (1).  
   By Theorem \ref{thm:MultiEdgeCreate}, $\pi_F \cup \pi_G$ is a minimum flip plan that starts from the equilateral triangulation $T$ and forces $F \cup G$ to become edges.  
   Also, by Lemma \ref{lem:flp_target_minimum}, the target triangulation of $\pi_F$ is the minimum triangulation $MT(S,F)$.  
   Hence, $\pi_G \setminus \pi_F$ is a flip plan that starts from the minimum triangulation $MT(S,F)$ and forces $G$ to become edges.  
   Additionally, since the line-segments of any point-pair in $F$ and any point-pair in $G$ do not intersect, no flip in $\pi_G \setminus \pi_F$ removes a point-pair in $F$, by Lemma \ref{lem:farey_parallelogram_intersect}.  
   Consequently, $\pi_G \setminus \pi_F$ is constrained by $F$.  
   Lastly, if $\pi_G \setminus \pi_F$ is not a minimum flip plan, then let $\pi$ be such a minimum flip plan.  
   Clearly, $\pi_F || \pi$ is a flip plan that starts from $T$ and forces $F \cup G$ to become edges.  
   Furthermore, it contains fewer flips than $\pi_F \cup \pi_G$, contradicting Theorem \ref{thm:MultiEdgeCreate}.
   Therefore, Statement (1) is proved.  
   Additionally, if some other minimum flip plan contains a different set of flips, then we get a similar contradiction.  
   
   Next, we prove Statement (2).  
   By Lemma \ref{lem:flp_target_minimum}, the target triangulation of $\pi_F \cup \pi_G$ is the minimum triangulation $MT(S,F \cup G)$.  
   Hence, $\pi_G \setminus \pi_F$ is clearly a flip plan between $MT(S,F)$ and $MT(S,F \cup G)$, and it is constrained by $F$ as shown above.  
   If it is not also a minimum flip plan, then let $\pi'$ be such a minimum flip plan.  
   By definition, $MT(S,F \cup G)$ contains $F \cup G$ as edges, so we get the same contradiction as above by considering the flip plan $\pi_F || \pi'$.  
   Therefore, Statement (2) is true.  
   
   Finally, if Statement (3) is false, then it is easy to argue that the minimum flip plan starting from an equilateral triangulation and forcing $G$ to become edges is not unique, contradicting Theorem \ref{thm:MultiEdgeCreate}.  
   Thus, Statement (3) is true, and so the lemma  is proved.  
\end{proof}

\end{document}